\documentstyle[aps,prl,preprint] {revtex}
\newcommand{\beq}{\begin{equation}} \newcommand{\eeq}{\end{equation}}
\newcommand{\bea}{\begin{eqnarray}} \newcommand{\eea}{\end{eqnarray}}
\newcommand{\bear}{\begin{eqnarray*}} \newcommand{\eear}{\end{eqnarray*}}
\newcommand{\lb}{\label} 
\newcommand{\rf}[1]{(\ref{#1})}

\begin{document}


\title
{Exact solutions of exactly integrable quantum chains by a matrix product
ansatz}

\author{Francisco C. Alcaraz  and Matheus J. Lazo }

\address {Universidade de S\~ao Paulo, Instituto de F\'{\i}sica de S\~ao 
Carlos, Caixa Postal 369, \\ 13560-590 S\~ao Carlos, S\~ao Paulo, Brazil}

\date{\today}

\maketitle


\begin{abstract}

Most of the exact solutions of quantum one-dimensional Hamiltonians are
obtained thanks to the success of the Bethe ansatz on its several formulations.
According to this ansatz the amplitudes of the
eigenfunctions of the Hamiltonian are given by a
sum of permutations of appropriate plane waves. In this paper, alternatively,
we present a
matrix product ansatz that asserts that those amplitudes are given in terms of
a matrix product. The eigenvalue equation for the Hamiltonian define the
algebraic properties of the matrices defining the amplitudes. The existence
of a consistent algebra imply the exact integrability of the model. The matrix
product ansatz we propose allow an unified and simple formulation of several
exact integrable Hamiltonians. In order to introduce and illustrate this
ansatz we present the exact solutions of several quantum chains with one and
two global conservation laws and periodic boundaries such as the XXZ chain,
spin-1 Fateev-Zamolodchikov model, Izergin-Korepin model, Sutherland model, t-J
model, Hubbard model, etc. Formulation of the matrix product ansatz for quantum
chains with open ends is also possible. As an illustration we present the exact
solution of an extended XXZ chain with $z$-magnetic fields at the surface and
arbitrary hard-core exclusion among the spins.

\end{abstract}

\section{Introduction}
        The Bethe  ansatz \cite{bethe} and its generalizations merged along the years as a quite efficient and powerful tool for the exact solution of the eigenspectrum of a great variety of one-dimensional quantum chains and two-dimensional transfer matrices (see e.g. \cite{baxter}-\cite{revschlo} for reviews). According to this  ansatz the amplitudes of the wavefunctions are expressed by a nonlinear combination  of properly defined plane waves. Although the question of completeness of the Bethe solutions is in general an open and difficult problem \cite{baxter2,nepome1}, a quantum Hamiltonian is considered exactly integrable if, in the thermodynamic limit, an infinite number of its eigenfunctions are given by the Bethe  ansatz.
  
        On the other hand along the last two decades \cite{affleck}-\cite{klumper}, several models were discovered having a ground state wavefunction exactly known and obtained though an {\it ansatz} known as the matrix product {\it ansatz}. Differently from the Bethe {\it ansatz}, this approach only gives the ground state eigenfunction, whose amplitudes are expressed in terms of a product of matrices, or more generally in terms of a product of generators of quadratic algebras \cite{alcritda}. In a distinct context a matrix product {\it ansatz} has also been applied quite successfully to the exact solution of the stationary distribution of probabilities of some one-dimensional stochastic models \cite{derr1}. The similarity between the master equation describing the time fluctuations of these models and the Schr\"odinger equation in Euclidean times enables to identify an associated ``quantum'' Hamiltonian to these stochastic models. The simplest example is the problem of asymmetric diffusion of hard-core particles on the one-dimensional lattice (see \cite{derr2}-\cite{alcrit1} for reviews). The time evolution operator governing the time fluctuations of this last model coincides with the exact integrable anisotropic Heisenberg chain, or XXZ chain, on its ferromagnetic regime and appropriate boundary fields \cite{alcrit1}. The related quantum chains in general are not exact integrable but have their ground state eigenfunctions given in terms of a matrix product {\it ansatz}. The matrix product {\it ansatz}, although providing only stationary properties of some stochastic system produced interesting results in a quite variety of problem including interface growth \cite{krug}, boundary induced phase transitions \cite{derr1}, \cite{derdomany}-\cite{derrevans} the dynamics of shocks \cite{derlebov} or traffic flow \cite{nagel}.

        More recently an interesting development of the matrix product  ansatz, named dynamical matrix product  ansatz \cite{stinchshutz}, merged in the area of stochastic one-dimensional systems. Models satisfying this  ansatz,  distinctly from the previous formulated matrix product  ansatz, have their probability distributions, at arbitrary times, given in terms of a product of matrices, which are now time dependent. The dynamical matrix product  ansatz was shown originally to be valid in the problem of asymmetric diffusion of particles on the lattice \cite{stinchshutz,sasamowada1}. More recently \cite{popkov,popkov2} the validity of this  ansatz was also confirmed in the exact integrable manifold of the asymmetric diffusion of particles with two kinds of particles. The validity of this  ansatz for such integrable system induces us to expect that all the quantum chains, related or not to the stochastic systems, that are solvable though the Bethe  ansatz may also be solvable by an appropriate matrix product  ansatz. We expect that all the components of an arbitrary eigenfunction of an exact integrable quantum chain, that are normally given in terms of a combination of plane waves, can also be expressed in terms of a product of matrices satisfying algebraic properties that ensure the exact integrability of the model.

        In this paper we are going to show the validity of this conjecture for a huge family of exact integrable quantum Hamiltonians, by showing how to formulate their solutions in terms of a matrix product  ansatz. A brief summary of some of our results has been announced in \cite{letterJPA}. We are going to present explicit examples of a matrix product  ansatz formulation for models having one or two global conservations. Examples of models with one global conservation law (U($1$) symmetric) include the XXZ quantum chain \cite{yang}, the spin-$1$ Fateev-Zamolodchikov model \cite{fateev}, the Izergin-Korepin \cite{izergin} and the solvable spin-$1$ model considered in \cite{alcbar1}. Among the models with two global conservation laws (U($1$)$\otimes$U($1$) symmetric) we present the matrix product  ansatz for the models where, in its formulation in terms of particles, no double occupancy of sites is allowed, like the anisotropic spin-$1$ Sutherland model \cite{sutherland}, the  spin-$1$ Perk-Schultz models \cite{perkshultz},  the fermionic supersymmetric t-J model \cite{schlo}, and also for the models with double occupancy like the spin-$3/2$ Perk-Schultz model \cite{perkshultz}, the Essler-Korepin-Schoutens model \cite{essler2}, the Hubbard \cite{lieb} model and the two-parameter integrable model introduced in \cite{alcbar2}.

        The  ansatz we produce enable us to formulate in a simple and unified way generalized integrable models where the quantum spins have now hard-core interactions that exclude the occupation of two spins at neighbouring sites.  Although most of our results will be presented for periodic chains they can also be extended for non periodic but integrable chains. As an example of  a non periodic chain we present the solution, in terms of a matrix product  ansatz, for the exact integrable XXZ chain with boundary $z$-magnetic fields \cite{ab3q}.

        The layout of the paper is as follows. In sections $2$ and $3$ we consider models with a single global conservation law. In section $2$ we introduce and solve through our matrix product  ansatz a generalization of the standard XXZ chain, where the up spins have a hard core interaction of a given, but arbitrary, range in terms of units of lattice spacing. In section $3$ models with spin $1$ are considered. The matrix product  ansatz is formulated for the spin-$1$ Fateev-Zamolodchikov model, the Izergin-Korepin model, as well as for the spin-$1$ model introduced in \cite{alcbar1}. All these solutions are obtained through a single matrix product  ansatz, whose matrices satisfy for each model distinct algebraic relations. In section $4$ and $5$ we consider models with U($1$)$\otimes$U($1$) symmetry, thus exhibiting two conserved global quantities. In section $4$ we consider the models of spin-$1$ in this class. Those models are the anisotropic Sutherland model, or the SU($3$) Perk-Schultz model and the supersymmetric t-J model. In section $4$ we present a general solution that includes the spin-$3/2$ Perk-Schultz model, the Essler-Korepin-Schoutens model, the Hubbard model as well the general exact integrable two-parameter model presented in \cite{alcbar2}. Again the solution of all these models are presented through an unified matrix product  ansatz. All the solutions presented in the previous sections  are derived for quantum chain defined on lattices with toroidal boundary conditions, in section $6$ we show how to extend our solutions for the case of the XXZ chain in an open lattice with a $z$-magnetic field at the end points of the lattice. Finally in section $7$ we close our paper with some conclusions and comments.
       


\section{Generalized XXZ chains}

	As a first application of our matrix product  ansatz we are going to present, in this section, the exact solution of the anisotropic Heisenberg model or XXZ chain. In order to show the powerfulness of our  ansatz, instead of solving the standard XXZ chain we are going to solve a generalization of this quantum Hamiltonian where arbitrary excluded volume are considered among the up spins ($\sigma^z$-basis). We consider a generalized XXZ chain where any two up spins, due to hard-core interactions, are not allowed to occupy lattice sites at distances smaller than $s$ ($s=1,2,\ldots$), in units of lattice spacing. Unlike the down spins, that have only on site hard-core exclusions, the up spins behave as they would have an effective size $s$. The generalized XXZ Hamiltonian we consider, in a lattice with $L$ sites, is given by 
\beq
H_s=-{\cal P}_s \sum_{i=1}^{L}\frac{1}{2}(\sigma_{i}^{x}\sigma_{i+1}^{x}+\sigma_{i}^{y}\sigma_{i+1}^{y}+
\Delta\sigma_{i}^{z}\sigma_{i+s}^{z}){\cal P}_s,
\lb{e1}
\eeq
where $\sigma^x$, $\sigma^y$, $\sigma^z$ are spin-$1/2$ Pauli matrices, $\Delta$ is the anisotropy and the projector ${\cal P}_s$ projects out from the associated Hilbert space the configurations where any two up spins are at distance smaller than $s$. In the particular case where $s=1$ the projector ${\cal P}_s$ is the 
unity operator and we recover the standard XXZ chain. The Hamiltonian \rf{e1}, in terms of raising and lowering spin-$1/2$ operators $\sigma^{\pm}=(\sigma^x{\pm}i\sigma^y)/2$ is a particular case of the general Hamiltonian
\beq
H_s=-{\cal P}_s\sum_{i=1}^{L}\left[\epsilon_{+}\sigma_{i}^{-}\sigma_{i+1}^{+}+\epsilon_{-}\sigma_{i}^{+}\sigma_{i+1}^{-}+
\frac{\Delta}{2}(\sigma_{i}^{z}\sigma_{i+s}^{z}-1)\right]{\cal P}_s
\lb{e2}
\eeq
where $\epsilon_+=\epsilon_-=1$ and a harmless constant ($+L\Delta/2$) was added. Without any additional difficulty we are going to consider the solution of \rf{e2}, for general values of $\epsilon_+$, $\epsilon_-$ and $\Delta$. If we interpret the up spins as particles and the down spins as vacant sites the Hamiltonian \rf{e2} with the choice $\epsilon_+ +\epsilon_- =1=-2\Delta$ coincides, apart from a harmless constant, with the time evolution operator of the asymmetric diffusion problem (asymmetry $\epsilon_+/\epsilon_-$) of particles with size $s$ on the lattice \cite{PRE}. We are going to consider \rf{e2} with the periodic boundary condition
\beq
\sigma^{\pm}_{L+1}=\sigma^{\pm}_1, \;\;\; \sigma^z_{L+1}=\sigma^z_1.
\lb{e3}
\eeq
In section $6$ we are going to consider the case of open boundary conditions.

        The Hamiltonian \rf{e2} with the boundary condition \rf{e3} has a U($1$)$\otimes$Z$_L$ symmetry due to its commutation with the total spin operator $S^z=\sum_{i=1}^L\sigma^z$ and the spatial translation operator $\hat{T}=e^{i\hat{P}}$ on the lattice. Consequently the Hilbert space associated to \rf{e2} can be separated into block disjoint sectors  labelled by the number $n$ of up spins ($n=0,1,\ldots,L$) and the momentum eigenvalues $P$ ($P=\frac{2\pi}{L}l, l=0,1,\ldots,L-1$).

        The  ansatz we propose asserts that any eigenfunction $|\Psi_{n,P}\rangle$ of \rf{e2} in the sector with $n$ up spins and momentum $P$, will have its components given by the trace of the matrix product
\bea
|\Psi_{n,P}\rangle &=& \sum_{\{x_1,\ldots,x_n\}}^*f(x_1,\ldots,x_n)|x_1,\ldots,x_n\rangle, \nonumber\\
f(x_1,\ldots,x_n) &=& \mbox{Tr}\left[E^{x_1-1}A^{(s)}E^{x_2-x_1-1}A^{(s)}\cdots E^{x_n-x_{n-1}-1}A^{(s)}E^{L-x_n}\Omega_P\right].
\lb{e4}
\eea
The configuration in $\sigma^z$-basis where the up spins are located at ($x_1,\ldots,x_n$) are denoted by the ket $|x_1,\ldots,x_n\rangle$ and the symbol ($*$) in the sum denotes the restriction to the sets satisfying the hard-core exclusion due to the effective size $s$ of the up spins, i.e.,
\beq
x_{i+1}\ge x_i+s, \;\;\; i=1,\ldots,n-1, \;\;\; x_1\ge 1, \;\;\; s\le x_n-x_1\le L-s.
\lb{e5b}
\eeq
Differently from the standard Bethe  ansatz where $f(x_1,\ldots,x_n)$ is given by the combination of plane waves now it is given by the trace of a product of matrices. Actually $E$, $A^{(s)}$ and $\Omega_P$ are abstract objects with appropriate commutation relations. These operators in order to provide well defined amplitudes for the eigenfunctions should have an associative product, and as usual in the literature, we call them simply as matrices. The matrices $E$ and $A^{(s)}$ are associated to the down and up spins representing the ket configuration. The superscript $s$ is just to remember that the up spins have an effective size $s$. The matrix $\Omega_P$ in \rf{e4} is introduced in order to fix the momentum $P$ of the eigenfunction $|\Psi_{n,P}\rangle$. This is accomplished by imposing the algebraic relations
\footnote{In the general case of twisted boundary conditions
$\sigma_{L+1}^{\pm} = e^{\pm i\Phi}\sigma_1^{\pm}$, 
$\sigma_{L+1}^z = \sigma_1^z$ the  ansatz also works with an appropriate
generalization of \rf{e5}.}
\beq
E\Omega_P=e^{-iP}\Omega_PE, \;\;\; A^{(s)}\Omega_P=e^{-iP}\Omega_PA^{(s)}
\lb{e5}
\eeq
since from \rf{e4}, eigenfunctions of momentum $P$ should have the ratio of amplitudes
\beq
\frac{f(x_1,\ldots,x_n)}{f(x_1+m,\ldots,x_n+m)}=e^{-iPm} \;\;\; (m=0,1,\ldots,L-1).
\lb{e6}
\eeq
The matrix product  ansatz for $|\Psi_{n,P}\rangle$ will work if we are able to obtain consistent algebraic relations among $E$ and $A^{(s)}$ that solve the eigenvalue equation
\beq
H_s|\Psi_{n,P}\rangle=\varepsilon_n|\Psi_{n,P}\rangle.
\lb{e7}
\eeq

        As is customary, before considering the case of general values of $n$ let us consider the eigensectors with $n=1$ and $n=2$ up spins.

{\it {\bf  n = 1.}}
For one up spin the eigenvalue equation \rf{e7} give us for the amplitudes with a single spin at $x_1=1,\ldots,L$
\bea
\varepsilon_1\mbox{Tr}\left[E^{x_1-1}A^{(s)}E^{L-x_1}\Omega_P\right]&=&-\epsilon_+\mbox{Tr}\left[E^{x_1-2}A^{(s)}E^{L-x_1+1}\Omega_P\right]-\epsilon_-\mbox{Tr}\left[E^{x_1}A^{(s)}E^{L-x_1-1}\Omega_P\right]\nonumber\\
&&+\Delta\mbox{Tr}\left[E^{x_1-1}A^{(s)}E^{L-x_1}\Omega_P\right].
\lb{e8}
\eea
The cyclic property of the trace and relation \rf{e5} fix the eigenenergies
\beq
\varepsilon_1=-\left(\epsilon_+e^{-iP}+\epsilon_-e^{iP} - \Delta \right),
\lb{e9}
\eeq
where $P=\frac{2{\pi}l}{L}$ ($l=0,1,\ldots,L-1$) is the momentum of the state.
An alternative solution of \rf{e8} that will be simple to generalize for arbitrary values of $n$ is obtained by replacing 
\beq
A^{(s)}=A^{(s)}_kE^{2-s},
\lb{e10}
\eeq
where now $A^{(s)}_k$ is a spectral parameter dependent matrix obeying the following commutation relation with the matrix $E$:
\beq
EA^{(s)}_k=e^{ik}A^{(s)}_kE.
\lb{e11}
\eeq
Inserting \rf{e10} in \rf{e8} and using \rf{e11} we obtain 
\beq
\varepsilon_1=\varepsilon(k)=-\left(\epsilon_+e^{-ik}+\epsilon_-e^{ik}- 
\Delta\right),
\lb{e12}
\eeq
and comparing \rf{e9} with \rf{e12} we fix the spectral parameter $k$ as the momentum of the $1$-particle eigenfunction $|\Psi_{1,P}\rangle$, i. e., $k=P=\frac{2{\pi}l}{L}$ ($l=0,1,\ldots,L-1$). 

{\it {\bf  n = 2.}}
The eigenvalue equation \rf{e7} gives for the amplitudes of $|\Psi_{2,P}\rangle$ two types of relations depending if the two up spins located at $x_1$ and $x_2$ are at the closest position or not. The amplitudes corresponding to the kets $|x_1,x_2\rangle$ where $x_2>x_1+s$ give us the relation 
\bea
&&\varepsilon_2\mbox{Tr}\left[E^{x_1-1}A^{(s)}E^{x_2-x_1-1}A^{(s)}E^{L-x_2}\Omega_P\right]=-\epsilon_+\mbox{Tr}\left[E^{x_1-2}A^{(s)}E^{x_2-x_1}A^{(s)}E^{L-x_2}\Omega_P\right]\nonumber\\
&&-\epsilon_-\mbox{Tr}\left[E^{x_1}A^{(s)}E^{x_2-x_1-2}A^{(s)}E^{L-x_2}\Omega_P\right]-\epsilon_+\mbox{Tr}\left[E^{x_1-1}A^{(s)}E^{x_2-x_1-2}A^{(s)}E^{L-x_2+1}\Omega_P\right]\nonumber\\
&&-\epsilon_-\mbox{Tr}\left[E^{x_1-1}A^{(s)}E^{x_2-x_1}A^{(s)}E^{L-x_2-1}\Omega_P\right]+2\Delta\mbox{Tr}\left[E^{x_1-1}A^{(s)}E^{x_2-x_1-1}A^{(s)}E^{L-x_2}\Omega_P\right].
\lb{e13}
\eea
A possible and convenient way to solve this equation is obtained by a generalization of \rf{e10} and \rf{e11}. We identify $A^{(s)}$ as composed by two spectral parameter dependent new matrices $A^{(s)}_{k_1}$ and $A^{(s)}_{k_2}$, i. e.,
\beq
A^{(s)}=\sum_{i=1}^2A^{(s)}_{k_i}E^{2-s},
\lb{e14}
\eeq
that satisfy the commutation relations
\beq
EA^{(s)}_{k_j}=e^{ik_j}A^{(s)}_{k_j}E, \;\;\; \left(A^{(s)}_{k_j}\right)^2=0,  \;\;\; j=1,2.
\lb{e15}
\eeq
Inserting \rf{e14} in \rf{e13} and using \rf{e15} we obtain
\beq
\varepsilon_2=\varepsilon(k_1)+\varepsilon(k_2),
\lb{e16}
\eeq
where $\varepsilon(k)$ is given by \rf{e12}. The relation \rf{e5} give us the commutation of the new matrices $A^{(s)}_{k_i}$ with $\Omega_P$, i. e.,
\beq
A^{(s)}_{k_j}\Omega_P=e^{iP(1-s)}\Omega_PA^{(s)}_{k_j} \;\;\; j=1,2.
\lb{e17}
\eeq
Comparing as in \rf{e6} the amplitudes of the configurations $|x_1,x_2\rangle$ and $|x_1+m,x_2+m\rangle$ and exploring the cyclic property of the trace we obtain
\beq
P=k_1+k_2.
\lb{e18}
\eeq

	Up to now we have at our disposal,  for solving the eigenvalue equation, the commutation relation of $A^{(s)}_{k_1}$ and $A^{(s)}_{k_2}$ as well the spectral parameters $k_1$ and $k_2$,  that might be in general complex numbers. These commutations relations are going to be fixed by the application of the eigenvalue equation \rf{e7} to the components of the configurations $|x_1,x_2\rangle$ where the up spins are at the ``matching'' condition $x_2=x_1+s$:
\bea
&&\varepsilon_2\mbox{Tr}\left[E^{x_1-1}A^{(s)}E^{s-1}A^{(s)}E^{L-x_1-s}\Omega_P\right]=-\epsilon_+\mbox{Tr}\left[E^{x_1-2}A^{(s)}E^{s}A^{(s)}E^{L-x_1-s}\Omega_P\right]\nonumber\\
&&-\epsilon_-\mbox{Tr}\left[E^{x_1-1}A^{(s)}E^{s}A^{(s)}E^{L-x_1-s-1}\Omega_P\right]+\Delta\mbox{Tr}\left[E^{x_1-1}A^{(s)}E^{s-1}A^{(s)}E^{L-x_1-s}\Omega_P\right].
\lb{e19}
\eea
Using \rf{e14}, \rf{e16} and the commutation relations \rf{e15} in the last expression one obtains
\beq
\sum_{j,l=1}^2\left[\epsilon_-+\epsilon_+e^{-i(k_j+k_l)}-{\Delta}e^{-ik_j}\right]A^{(s)}_{k_j}A^{(s)}_{k_l}=0.
\lb{e20}
\eeq
This imply that the matrices $A^{(s)}_{k_j}$ ($j=1,2$) should obey 
\beq
A^{(s)}_{k_j}A^{(s)}_{k_l}=S(k_j,k_l)A^{(s)}_{k_l}A^{(s)}_{k_j}, \;\;\; (l{\neq}j), \;\;\; \left(A^{(s)}_{k_j}\right)^2=0, \;\;\; (j,l=1,2),
\lb{e21}
\eeq
where
\beq
S(k_j,k_l)=-\frac{\epsilon_++\epsilon_-e^{i(k_j+k_l)}-{\Delta}e^{ik_j}}{\epsilon_++\epsilon_-e^{i(k_j+k_l)}-{\Delta}e^{ik_l}}.
\lb{e22}
\eeq
This last relation in the context of $(1+1)$-dimensional field theory is know as the Zamolodchikov algebra \cite{zamo1,zamo2}.
The complex spectral parameters $k_j$ ($j=1,2$) that are still free up to now are fixed by imposing that the ratio of two components $f(x_1,x_2)/f(x_1',x_2')$ should be uniquely related. The cyclic property of the trace with the algebraic relations \rf{e15}, \rf{e16} and \rf{e21} give us 
\bea
\mbox{Tr}\left[ A_{k_l}^{(s)}A_{k_j}^{(s)}E^{L-2s+2}\Omega_P\right] &=& e^{-i(L-2s+2)k_j}\mbox{Tr}\left[ A_{k_l}^{(s)}E^{L-2s+2}A_{k_j}^{(s)}\Omega_P\right] \nonumber \\ 
&=& e^{-ik_j L} e^{i2k_j(s-1)}e^{-iP(s-1)} S(k_j, k_l)\mbox{Tr}\left[ A_{k_l}^{(s)}A_{k_j}^{(s)}E^{L-2s+2}\Omega_P\right],
\lb{e23}
\eea
or equivalently, since $P=k_1+k_2$,
\beq
e^{ik_j L} = S(k_j, k_l) \left( \frac{e^{ik_j}}{e^{ik_l}} \right)^{s-1},\;\;\; j=1,2\;\;\; (j \neq l).
\lb{e24}
\eeq
The energies $\varepsilon_2$ and momentum $P$ of $|\Psi_{2,P}\rangle$ are obtained by inserting the solutions of \rf{e24} into \rf{e16} and \rf{e18} respectively.

{\it {\bf General n.}}
The previous calculation can be easily extended for arbitrary values of the number $n$ of up spins. The eigenvalue equation \rf{e7} when applied to the amplitudes of $|\Psi_{n,P}\rangle$ corresponding to the configurations where all the $n$ spins are at distances larger than the excluded volume $s$, gives a generalization of \rf{e13}:
\bea
\varepsilon_n\mbox{Tr}\left[ \cdots E^{x_i-x_{i-1}-1}A^{(s)}E^{x_{i+1}-x_i-1}A^{(s)} \cdots A^{(s)} E^{L-x_n}\Omega_P\right] = \nonumber\\  
- \sum_{i=1}^n \{ \epsilon_+\mbox{Tr}\left[ \cdots E^{x_i-x_{i-1}-2}A^{(s)}E^{x_{i+1}-x_i}A^{(s)} \cdots A^{(s)}E^{L-x_n}\Omega_P \right] \nonumber \\
+ \epsilon_-\mbox{Tr}\left[ \cdots E^{x_i-x_{i-1}-1}A^{(s)} E^{x_{i+1}-x_i-2}A^{(s)} \cdots A{(s)}E^{L-x_n+1}\Omega_P\right] \nonumber \\
-\Delta\mbox{Tr}\left[ \cdots E^{x_i-x_{i-1}-1}A^{(s)}E^{x_{i+i}-x_i-1}A^{(s)} \cdots A^{(s)}E^{L-x_n}\Omega_P \right] \}.
\lb{e25}
\eea
The solution of this last equation\footnote{The most general relation $A^{(s)}=\sum_{j=1}^nE^{\alpha}A_{k_j}^{(s)}E^{\beta}$ could be used. However \rf{e26} is more convenient since otherwise the structure constants $S(k_i,k_j)$ in \rf{e22} will depend on the size $s$.} is obtained by identifying the $A^{(s)}$ matrix as a combination of $n$ spectral parameter dependent matrices $\{A_{k_j}; j =1,\ldots,n\}$, i. e.,
\beq
A^{(s)}=\sum_{j=1}^n A_{k_j}^{(s)}E^{2-s},
\lb{e26}
\eeq
with the commutation relations with the matrices $E$ and $\Omega_P$ given by 
\beq
E A_{k_j}^{(s)} = e^{ik_j} A_{k_j}^{(s)}E, \;\;\; A_{k_j}^{(s)} \Omega_P = e^{iP(1-s)} \Omega_P A_{k_j}^{(s)} \;\;\; (j=1, \ldots ,n).
\lb{e27}
\eeq
The energy and momentum are obtained by inserting \rf{e26} into \rf{e25} and using \rf{e27},  and are given by 
\beq
\varepsilon_n = \sum_{j=1}^n \varepsilon(k_j),\;\;\; P = \sum_{j=1}^n k_j,
\lb{e28}
\eeq
respectively. The eigenvalue equation \rf{e7} gives for the configuration where a pair of up spins are at the ``colliding'' positions $x_{i+1}=x_i+s$ a relation that coincides with \rf{e20}-\rf{e22} but with $j,l=1,\ldots,n$. The relation coming from the configuration where three particles are at the colliding position $x_{i+1}=x_i+s$, $x_{i+2}=x_{i+1}+s$ is given by
\bea
&&\sum_{j,l=1}^n\left(\epsilon_-+\epsilon_+e^{-i(k_j+k_l)}-\Delta e^{-ik_j}\right)A_{k_j}^{(s)}A_{k_l}^{(s)}\sum_{t=1}^nA_{k_t}^{(s)}e^{ik_t} \nonumber \\
&&+\sum_{j=1}^ne^{-ik_j}A_{k_j}^{(s)}\sum_{j,l=1}^ne^{i(k_l+k_t)}\left(\epsilon_-+\epsilon_+e^{-i(k_l+k_t)}-\Delta e^{-ik_l}\right)A_{k_l}^{(s)}A_{k_t}^{(s)}=0,
\lb{e29}
\eea
being a  consequence of the relation \rf{e20}. Similarly the amplitudes with arbitrary number of particles at colliding position will be automatically satisfied if the matrices $E$, $A_{k_j}^{(s)}$ ($j=1,\ldots,n$) and $\Omega_P$ obey the algebraic relations \rf{e21} and \rf{e27}. The associativity of the algebra provide a well defined value for any product of matrices and follows from the fact that the structure constants $S(k_i,k_j)$ in \rf{e21}-\rf{e22} are c-numbers with the property $S(k_i,k_j)S(k_j,k_i)=1$ ($i,j=1,\ldots,n$).

        The cyclic invariance of the trace in \rf{e4} will fix the complex spectral parameters $k_j$ ($j=1,\ldots,n$), providing  a well defined value for the components of $|\Psi_{n,P}\rangle$, i. e., 
\bea
&&\mbox{Tr}\left[A_{k_1}^{(s)}\cdots A_{k_j}^{(s)}\cdots A_{k_n}^{(s)}E^{L-n(s-1)}\Omega_P \right] \nonumber \\
&&=e^{-ik_j[L-n(s-1)]}e^{iP(1-s)}\prod_{l=j+1}^nS(k_j,k_l)\mbox{Tr}\left[A_{k_j}^{(s)}A_{k_1}^{(s)}\cdots A_{k_{j-1}}^{(s)}A_{k_{j+1}}^{(s)}\cdots A_{k_n}^{(s)}E^{L-n(s-1)}\Omega_P \right]\nonumber \\
&&=e^{-ik_j[L-n(s-1)]}e^{iP(1-s)}\prod_{l=1,l\neq j}^nS(k_j,k_l)\mbox{Tr}\left[A_{k_1}^{(s)}\cdots A_{k_j}^{(s)}\cdots A_{k_n}^{(s)}E^{L-n(s-1)}\Omega_P \right], \;\;\; j=1,\ldots,n, \nonumber \\
\lb{e30}
\eea
or equivalently
\beq
e^{ik_jL}=(-1)^n \prod_{l=1}^n \left( \frac{e^{ik_j}}{e^{ik_l}} \right)^{s-1} \frac{\epsilon_+ + \epsilon_-e^{i(k_j +k_l)} -\Delta e^{ik_j}}{\epsilon_+ + \epsilon_-e^{i(k_j +k_l)} -\Delta e^{ik_l}}, \;\;\; j=1,\ldots,n.
\lb{e31}
\eeq
The acceptable set $\{k_j; j=1,\ldots,n  \}$ of spectral parameters defining the eigenvectors $|\Psi_{n,P}\rangle$ are the solution of \rf{e31} where $k_i \neq k_j$ ($i,j=1,\ldots,n$). Since $(A_{k}^{(s)})^2=0$, solutions of \rf{e31} with coinciding roots give us null states.

        The equation \rf{e31} coincides with the Bethe  ansatz equation derived for the model \rf{e2} through the standard Bethe  ansatz \cite{PRE}. The choice $\epsilon_+=\epsilon_-=1$ with $s=1$ gives the Bethe  ansatz equation of the standard XXZ chain \rf{e1} \cite{yang}. Moreover, using \rf{e26},   an arbitrary component $f(x_1,\ldots,x_n)$ of the eigenfunction $|\Psi_{n,P}\rangle$, given by the  ansatz \rf{e4}, can be written as 
\bea
&&f(x_1, \ldots, x_n) = \nonumber \\
&&\sum_{i_1=1}^n \sum_{i_2=1}^n \cdots \sum_{i_n=1}^n \mbox{Tr} \left[ E^{x_1-1}A_{k_{i_1}}^{(s)}E^{x_2-x_1+1-s}A_{k_{i_2}}^{(s)} \cdots E^{x_n-x_{n-1}+1-s}A_{k_{i_n}}^{(s)} E^{L-x_n+2-s}\Omega_P\right].
\lb{e32}
\eea
The commutation relations \rf{e27}  allow us to write
\bea
&&f(x_1, \ldots, x_n) = \nonumber \\
&&\sum_{i_1=1}^n \cdots \sum_{i_n=1}^n e^{i[k_{i_1}(x_1-1)+\cdots +k_{i_n}(x_n-1)]} \mbox{Tr} \left[ A_{k_{i_1}}^{(s)}E^{1-s}A_{k_{i_2}}^{(s)}E^{1-s} \cdots E^{1-s}A_{k_{i_n}}^{(s)} E^{L}\Omega_P\right].
\lb{e33}
\eea
Let us define the new matrices
\beq
\tilde{A}_{k_j}^{(s)}=A_{k_j}^{(s)}E^{1-s}\;\;\; (j=1,\ldots ,n).
\lb{e34}
\eeq
It is simple to verify, from \rf{e21}, that they satisfy, for $j,l=1,\ldots,n$, 
\beq
\tilde{A}_{k_j}^{(s)} \tilde{A}_{k_l}^{(s)} = \tilde{S}(k_j, k_l) \tilde{A}_{k_l}^{(s)} \tilde{A}_{k_j}^{(s)}, \;\;\; (j \neq l), \;\;\; \left( \tilde{A}_{k_j}^{(s)} \right)^2 = 0, 
\lb{e35}
\eeq
where
\beq
\tilde{S}(k_j, k_l)=S(k_j, k_l)\left( \frac{e^{ik_j}}{e^{ik_l}} \right)^{s-1}.
\lb{e36}
\eeq
In terms of these new matrices, and exploring the fact that $( \tilde{A}_{k_j}^{(s)})^2 = 0$ we can write
\beq
f(x_1, \ldots, x_n) = \sum_{p_1, \ldots, p_n} e^{i[k_{p_1}(x_1-1)+ \cdots +k_{p_n}(x_n-1)]} \mbox{Tr} \left[ \tilde{A}_{k_{p_1}}^{(s)}\cdots \tilde{A}_{k_{p_n}}^{(s)} E^{L}\Omega_P\right],
\lb{e37}
\eeq
where the sum is over the permutations $\{p_1, p_2, \ldots,p_n\}$ of the non repeated integers ($1,\ldots,n$). This last result show us that the amplitudes derived using the proposed matrix product  ansatz \rf{e4} is given by a combination of plane waves with complex wave numbers $\{k_i \}$, and reproduces the results previously obtained for the Hamiltonian \rf{e2} \cite{PRE} through the coordinate Bethe  ansatz.


\section{ Models of spin $1$ with one conservation law}

	We present in this section the appropriate matrix  ansatz solution for models of spin $1$ with a single global conservation law, like the XXZ chain of last section. Integrable models on this category are the Fateev-Zamolodchikov quantum chain \cite{fateev}, the Izergin-Korepin model \cite{izergin} and the spin-$1$ model introduced in \cite{alcbar1}.

        We start with a general spin-$1$ model with U($1$) symmetry and nearest neighbor interaction. Instead of writing this general model in terms of spin-$1$ Pauli matrices it is more convenient to write it in terms of the $3\times 3$ Weyl matrices $E^{l,m}$ ($l,m=0,1,2$), with $i,j$ elements $(E^{l,m})_{i,j}=\delta_{l,j}\delta_{m,j}$. At each lattice site we may have zero particles ($s_z=-1$), one particle ($s_z=0$) or two particles ($s_z=1$). The general Hamiltonian we consider, that conserves the number of particles, is given by
\bea
&& H_{s=1}^{U(1)}=-\sum_{i=1}^L h_{i,i+1}+L\Gamma_{0\;0}^{0\;0}, \nonumber \\
&& h_{i,i+1}=\sum_{k,l,m,n=0}^2 \Gamma_{m\;n}^{k\;l}E_i^{m,k}E_{i+1}^{n,l} \;\;\; (i=1,\ldots,L)
\lb{e38}
\eea
where periodic boundary conditions are imposed and in order to conserve the total number of particles $\Gamma_{m\;n}^{k\;l}=0$ if $k+l \neq m+n$.

	The total number of particles $n$ ($0,1,\ldots,2L$) and the momentum $P$ ($\frac{2\pi l}{L},\;l=0,\ldots,L-1$) are good quantum numbers for the general Hamiltonian \rf{e38}. The eigenfunctions on these eigensectors are given by
\beq
|\Psi_{n,P}\rangle=\sum_{\{x_1,\ldots,x_n \}}^* f(x_1,\ldots,x_n)|x_1,\ldots,x_n\rangle
\lb{e39}
\eeq
where ($x_1,\ldots,x_n$) are the coordinates of the particles and the symbol ($*$) in the sum means restriction to the sets with coordinates $\{x_{i+1}\ge x_i,\; x_{i+2}>x_i \}$.

	In order to formulate an appropriate matrix product  ansatz for \rf{e38} we associate, as in the previous section, the matrices $E$ and $A$ to the empty sites ($s_z=-1$) and the sites occupied by a single particle ($s_z=0$), respectively. The sites with double occupancy of particles ($s_z=1$) are associated with the matrix $BE^{-1}B$. Certainly the Hamiltonian \rf{e38} is not exact integrable for arbitrary values of $\{\Gamma_{m\;n}^{k\;l} \}$. The  ansatz we propose states that in the exact integrable manifold of \rf{e38} any eigenfunction $|\Psi_{n,P}\rangle$, in the sector with $n$ particles and momentum $P$ will have components given in terms of traces of a matrix product. The amplitudes corresponding to the configurations where there is no double occupancy are given by 
\beq
f(x_1,\ldots,x_n)=\mbox{Tr} \left[E^{x_1-1}AE^{x_2-x_1-1}A\cdots E^{x_n-x_{n-1}-1}AE^{L-x_n}\Omega_P \right]
\lb{e40}
\eeq
while if there exists a double occupancy at $x_{i+1}=x_i$ we have 
\bea
&&f(x_1,\ldots, x_{i-1},x_i,x_{i+1}, \ldots,x_n)= \nonumber \\
&&\mbox{Tr} \left[E^{x_1-1}A\cdots E^{x_i-x_{i-1}-1}BE^{-1}BE^{x_{i+2}-x_i-1}\cdots  E^{x_n-x_{n-1}-1}AE^{L-x_n}\Omega_P \right].
\lb{e41}
\eea
As in previous section the matrix $\Omega_P$ and the trace operation are introduced in order to fix the momentum $P=\frac{2\pi}{L}l$, ($l=0,1,\ldots,L-1$) of $|\Psi_{n,P}\rangle$, this is accomplished (see  \rf{e6}) by imposing
\beq
E\Omega_P=e^{-iP}\Omega_PE, \;\;\; A\Omega_P=e^{-iP}\Omega_PA, \;\;\; B\Omega_P=e^{-iP}\Omega_PB.
\lb{e42}
\eeq
The algebraic relations among the $E$, $A$ and $B$ operators will be obtained from the requirement that \rf{e38} with the  ansatz \rf{e39}-\rf{e41}  are solutions of the eigenvalue equation
\beq
H_{s=1}^{U(1)}|\Psi_{n,P}\rangle=\varepsilon_n|\Psi_{n,P}\rangle.
\lb{e43}
\eeq

	Let us consider initially the cases with a small number of particles. 

{\it {\bf  n = 1.}}
The eigenvalue equation gives for the amplitudes with a particle at position $x_1=1,\ldots,N$:
\bea
&&\varepsilon_1 \mbox{Tr} \left[E^{x_1-1}AE^{L-x_1}\Omega_P \right]=-\Gamma_{0\;1}^{1\;0}  \mbox{Tr} \left[E^{x_1-2}AE^{L-x_1+1}\Omega_P \right]-\Gamma_{1\;0}^{0\;1}  \mbox{Tr} \left[E^{x_1}AE^{L-x_1-1}\Omega_P \right]\nonumber \\
&& +\left(2\Gamma_{0\;0}^{0\;0}-\Gamma_{1\;0}^{1\;0}-\Gamma_{0\;1}^{0\;1} \right)\mbox{Tr} \left[E^{x_1-1}AE^{L-x_1}\Omega_P \right].
\lb{e44}
\eea
As in previous section a convenient solution is obtained by introducing the spectral parameter matrix $A_k$,
\beq
A=A_kE \;\;\; \mbox{with} \;\;\; EA_k=e^{ik}A_kE.
\lb{e45}
\eeq
Inserting \rf{e45} into \rf{e43} we obtain
\bea
&&\varepsilon_1=\varepsilon(k)=-\left(\Gamma_{0\;1}^{1\;0}e^{-ik}+\Gamma_{1\;0}^{0\;1}e^{ik}-2\Gamma_{0\;0}^{0\;0}+\Gamma_{1\;0}^{1\;0}+\Gamma_{0\;1}^{0\;1} \right),\nonumber \\
&&P=k=\frac{2\pi l}{L} \;\;\; (l=0,\ldots,L-1).
\lb{e46}
\eea

{\it {\bf  n = 2.}}
In this case the eigenvalue equation \rf{e42} produces relations that are distinct if $x_2>x_1+1$, $x_2=x_1+1$ or $x_2=x_1$. The relations coming from the amplitudes where the particles are not at the ``colliding'' positions $x_2=x_1$ or $x_2=x_1+1$ are just a straightforward generalization of \rf{e44}, for two particles, whose solution is obtained by identifying the matrix $A$ as composed by two spectral parameter dependent matrices $A_{k_1}$, $A_{k_2}$, i. e., 
\beq
A=\sum_{j=1}^nA_{k_j}E
\lb{e47}
\eeq
satisfying the commutations relations
\beq
EA_{k_j}=e^{ik_j}A_{k_j}E, \;\;\; \left(A_{k_j} \right)^2=0 \;\;\; j=1,\ldots,n.
\lb{e47p}
\eeq
The energy and momentum are given by
\beq
\varepsilon_n=\sum_{j=1}^n\varepsilon(k_j) \,\,\, P=\sum_{j=1}^nk_j,
\lb{e48}
\eeq
with $\varepsilon(k)$ given by \rf{e46}. The eigenvalue equation \rf{e43} gives for the components $f(x_1,x_1+1)$ the equation
\bea
&&\varepsilon_2 \mbox{Tr} \left[E^{x_1-1}AAE^{L-x_1-1}\Omega_P \right]=-\Gamma_{0\;1}^{1\;0}  \mbox{Tr} \left[E^{x_1-2}AEAE^{L-x_1-1}\Omega_P \right] \nonumber \\
&& -\Gamma_{1\;0}^{0\;1}  \mbox{Tr} \left[E^{x_1-1}AEAE^{L-x_1-2}\Omega_P \right]-\Gamma_{1\;1}^{2\;0}  \mbox{Tr} \left[E^{x_1-1}BE^{-1}BE^{L-x_1}\Omega_P \right]\nonumber \\
&&-\Gamma_{1\;1}^{0\;2}  \mbox{Tr} \left[E^{x_1}BE^{-1}BE^{L-x_1-1}\Omega_P \right]+\left(3\Gamma_{0\;0}^{0\;0}-\Gamma_{1\;0}^{1\;0}-\Gamma_{0\;1}^{0\;1}-\Gamma_{1\;1}^{1\;1} \right)\mbox{Tr} \left[E^{x_1-1}AAE^{L-x_1-1}\Omega_P \right]. \nonumber \\
\lb{e49}
\eea
while the components $f(x_1,x_1)$ gives
\bea
&&\varepsilon_2 \mbox{Tr} \left[E^{x_1-1}BE^{-1}BE^{L-x_1}\Omega_P \right]= \nonumber \\
&&-\Gamma_{0\;2}^{1\;1}  \mbox{Tr} \left[E^{x_1-2}AAE^{L-x_1}\Omega_P \right]-\Gamma_{2\;0}^{1\;1}  \mbox{Tr} \left[E^{x_1-1}AAE^{L-x_1-1}\Omega_P \right]\nonumber \\
&&-\Gamma_{0\;2}^{2\;0}  \mbox{Tr} \left[E^{x_1-2}BE^{-1}BE^{L-x_1+1}\Omega_P \right]-\Gamma_{2\;0}^{0\;2}  \mbox{Tr} \left[E^{x_1}BE^{-1}BE^{L-x_1-1}\Omega_P \right] \nonumber \\
&&+\left(2\Gamma_{0\;0}^{0\;0}-\Gamma_{2\;0}^{2\;0}-\Gamma_{0\;2}^{0\;2} \right)\mbox{Tr} \left[E^{x_1-1}BE^{-1}BE^{L-x_1}\Omega_P \right].
\lb{e50}
\eea
Both equations \rf{e49}, \rf{e50} are solved by using \rf{e47} and 
by writing  the matrix $B$ also as a combination of  two spectral parameter dependent matrices $B_{k_1}$, $B_{k_2}$, i. e.,
\beq
B=\sum_{j=1}^nB_{k_j}E \;\;\; \mbox{with} \;\;\; E B_{k_j} = e^{ik_j} B_{k_j}E, \;\;\; \left(B_{k_j} \right)^2=0 \;\;\; (j=1, \ldots ,n).
\lb{e51}
\eeq
Notice that the matrices $A_{k_j}$, $B_{k_j}$ have the same set $\{k_1,k_2 \}$ of undetermined complex spectral parameters. Inserting \rf{e47}-\rf{e48} and \rf{e51} in \rf{e49} and \rf{e50} we obtain
\beq
\sum_{j,l=1}^nN(k_j,k_l)\mbox{Tr} \left[E^{x_1}A_{k_j}A_{k_l}E^{L-x_1}\Omega_P \right]=\sum_{j,l=1}^nC_1(k_j,k_l)\mbox{Tr} \left[E^{x_1}B_{k_j}B_{k_l}E^{L-x_1}\Omega_P \right]
\lb{e52}
\eeq
and
\beq
\sum_{j,l=1}^nC_0(k_j,k_l)\mbox{Tr} \left[E^{x_1}B_{k_j}B_{k_l}E^{L-x_1}\Omega_P \right]=\sum_{j,l=1}^nC_2(k_j,k_l)e^{ik_l}\mbox{Tr} \left[E^{x_1}A_{k_j}A_{k_l}E^{L-x_1}\Omega_P \right]
\lb{e53}
\eeq
where
\bea
&&N(k_j,k_l)=\Gamma_{0\;1}^{1\;0}+\left(\Gamma_{1\;0}^{1\;0}+\Gamma_{0\;1}^{0\;1}-\Gamma_{0\;0}^{0\;0}-\Gamma_{1\;1}^{1\;1} \right)e^{ik_l}+\Gamma_{1\;0}^{0\;1}e^{i(k_j+k_l)} \nonumber \\
&&C_1(k_j,k_l)=\Gamma_{1\;1}^{2\;0}+\Gamma_{1\;1}^{0\;2}e^{i(k_j+k_l)} \nonumber \\
&&C_0(k_j,k_l)=\Gamma_{0\;1}^{1\;0}(e^{ik_l}+ e^{ik_j}) + \Gamma_{1\;0}^{0\;1}
(e^{ik_l} + e^{ik_j}) e^{i(k_l+k_j)} \nonumber \\ 
&& +\left( 2\Gamma_{1\;0}^{1\;0}+2\Gamma_{0\;1}^{0\;1}-\Gamma_{2\;0}^{2\;0}-\Gamma_{0\;2}^{0\;2}-2\Gamma_{0\;0}^{0\;0} \right)e^{i(k_j+k_l)}-\Gamma_{0\;2}^{2\;0}-\Gamma_{2\;0}^{0\;2}e^{i2(k_j+k_l)} \nonumber \\
&&C_2(k_j,k_l)=\Gamma_{0\;2}^{1\;1}+\Gamma_{2\;0}^{1\;1}e^{i(k_j+k_l)}
\lb{e54}
\eea
The relations \rf{e52} and \rf{e53} imply
\beq
N(k_j,k_l)A_{k_j}A_{k_l}+N(k_l,k_j)A_{k_l}A_{k_j}=C_1(k_j,k_l)\left(B_{k_j}B_{k_l}+B_{k_l}B_{k_j}\right)
\lb{e55}
\eeq
\beq
C_0(k_j,k_l)\left(B_{k_j}B_{k_l}+B_{k_l}B_{k_j}\right)=C_2(k_j,k_l)\left(e^{ik_l}A_{k_j}A_{k_l}+e^{ik_j}A_{k_l}A_{k_j}\right),
\lb{e56}
\eeq
with $k,l=1,\ldots,n$. Multiplying \rf{e56} by the symmetric function $C_1(k_j,k_l)$ and using \rf{e55}, the last equation gives the relation
\bea
&&A_{k_j}A_{k_l}=S(k_j,k_l)A_{k_l}A_{k_j} \;\;\; (j \neq l), \nonumber \\
&&S(k_j,k_l)=-\frac{C_0(k_j,k_l)N(k_l,k_j)-C_1(k_j,k_l)C_2(k_j,k_l)e^{ik_j}}{C_0(k_j,k_l)N(k_j,k_l)-C_1(k_j,k_l)C_2(k_j,k_l)e^{ik_l}}; \;\;\; j,l=1,2.
\lb{e57}
\eea
The spectral parameters $k_1$ and $k_2$ are going to be fixed by the cyclic property of the traces defining the amplitudes \rf{e40} and \rf{e41}. Since the $B$ matrices only appear in the combination
\beq
BE^{-1}B=\left(\sum_{j,l=1}^nB_{k_j}B_{k_l} \right)E.
\lb{e58}
\eeq
The use of \rf{e55} allow us to replace in any amplitude the $\{B_k \}$ matrices by the $\{A_k \}$ matrices. Then an arbitrary amplitude should be proportional to $\mbox{Tr} \left[A_{k_l}A_{k_j}E^{L}\Omega_P \right]$. The cyclic property of the trace and the commutation relations \rf{e42}, \rf{e47p} and \rf{e57} give us
\beq
e^{ik_jL}=S(k_j,k_l), \; \; j=1,2 \;\;\; (j \neq l)
\lb{e59}
\eeq
with $S(k_j,k_l)$ given by \rf{e57}. It is interesting to mention that we obtained for $n=2$ particles the integrability for arbitrary values of the coupling $\Gamma_{m\;n}^{k\;l}$ of the Hamiltonian \rf{e38}. Certainly this will be not the case for $n>2$ particles.

{\it {\bf  n = 3.}}
In this case we have several distinct types of relations for the amplitudes $f(x_1,x_2,x_3)$ in \rf{e40} and \rf{e41}. The eigenvalue equation when applied to the amplitudes where all the particle are not at ``colliding'' positions, i. e., $x_3>x_2+1>x_1+2$ give us a straightforward  generalization of \rf{e44}, for three particles, whose solution is obtained by identifying, as in \rf{e47}, the matrix $A$ as composed by three spectral parameter dependent matrices $A_{k_1}$, $A_{k_2}$, $A_{k_3}$. The energy and momentum are given by \rf{e48} with $n=3$. The components $f(x_1,x_2=x_1,x_3)$, $f(x_1,x_2=x_1+1,x_3)$ with $x_3>x_2+1$ and $f(x_1,x_2,x_3=x_2)$, $f(x_1,x_2,x_3=x_2+1)$ with $x_1<x_2-1$ give us generalizations of \rf{e55} and \rf{e56} for $n=3$. We have new relations when the eigenvalue equation is applied to the amplitudes corresponding to the $3$ particles at the colliding positions. This happens for $f(x,x,x+1)$
\bea
&&\varepsilon_3\mbox{Tr} \left[E^{x-1}BE^{-1}BAE^{L-x-1}\Omega_P \right]= \nonumber \\
&&-\Gamma_{0\;2}^{1\;1}\mbox{Tr} \left[E^{x-2}AAAE^{L-x-1}\Omega_P \right]-\Gamma_{1\;0}^{0\;1}\mbox{Tr} \left[E^{x-1}BE^{-1}BEAE^{L-x-2}\Omega_P \right] \nonumber \\
&&-\Gamma_{0\;2}^{2\;0}\mbox{Tr} \left[E^{x-2}BE^{-1}BEAE^{L-x-1}\Omega_P \right]-\Gamma_{2\;1}^{1\;2}\mbox{Tr} \left[E^{x-1}ABE^{-1}BE^{L-x-1}\Omega_P \right] \nonumber \\
&&+\left(3\Gamma_{0\;0}^{0\;0}-\Gamma_{0\;2}^{0\;2}-\Gamma_{2\;1}^{2\;1}-\Gamma_{1\;0}^{1\;0}\right)\mbox{Tr} \left[E^{x-1}BE^{-1}BAE^{L-x-1}\Omega_P \right]
\lb{e60}
\eea
and for amplitudes $f(x,x+1,x+1)$
\bea
&&\varepsilon_3\mbox{Tr} \left[E^{x-1}ABE^{-1}BE^{L-x-1}\Omega_P \right]= \nonumber \\
&&-\Gamma_{2\;0}^{1\;1}\mbox{Tr} \left[E^{x-1}AAAE^{L-x-2}\Omega_P \right]-\Gamma_{0\;1}^{1\;0}\mbox{Tr} \left[E^{x-2}AEBE^{-1}BE^{L-x-1}\Omega_P \right] \nonumber \\
&&-\Gamma_{2\;0}^{0\;2}\mbox{Tr} \left[E^{x-1}AEBE^{-1}BE^{L-x-2}\Omega_P \right]-\Gamma_{1\;2}^{2\;1}\mbox{Tr} \left[E^{x-1}BE^{-1}BAE^{L-x-1}\Omega_P \right] \nonumber \\
&&+\left(3\Gamma_{0\;0}^{0\;0}-\Gamma_{2\;0}^{2\;0}-\Gamma_{1\;2}^{1\;2}-\Gamma_{0\;1}^{0\;1}\right)\mbox{Tr} \left[E^{x-1}ABE^{-1}BE^{L-x-1}\Omega_P \right].
\lb{e61}
\eea
Inserting \rf{e47}, \rf{e48} and \rf{e51} with $n=3$ in \rf{e60} and \rf{e61} we obtain the new algebraic relations relating three spectral parameter matrices. Using \rf{e55} and \rf{e56} to simplify those expressions we obtain
\beq
\sum_{q,r,s=1}^3 \left(D_1(k_q,k_r,k_s)B_{k_q}B_{k_r}A_{k_s}+\Gamma_{2\;0}^{1\;1}e^{i(k_r+k_s)}A_{k_q}A_{k_r}A_{k_s}-\Gamma_{2\;1}^{1\;2}e^{i(k_r+k_s)}A_{k_q}B_{k_r}B_{k_s} \right)=0,
\lb{e62}
\eeq
\beq
\sum_{q,r,s=1}^3 \left(D_2(k_q,k_r,k_s)A_{k_q}B_{k_r}B_{k_s}+\Gamma_{0\;2}^{1\;1}e^{ik_s}A_{k_q}A_{k_r}A_{k_s}-\Gamma_{1\;2}^{2\;1}e^{ik_s}A_{k_q}B_{k_r}B_{k_s} \right)=0,
\lb{e63}
\eeq
where
\bea
&&D_1(k_1,k_2,k_3)=\Gamma_{0\;1}^{1\;0}+e^{i(k_1+k_2+k_3)}\Gamma_{2\;0}^{0\;2}-e^{ik_3}\left(\Gamma_{0\;0}^{0\;0}+\Gamma_{2\;1}^{2\;1}-\Gamma_{0\;1}^{0\;1}-\Gamma_{2\;0}^{2\;0}\right), \nonumber \\
&&D_2(k_1,k_2,k_3)=\Gamma_{0\;2}^{2\;0}+e^{i(k_1+k_2+k_3)}\Gamma_{1\;0}^{0\;1}-e^{i(k_2+k_3)}\left(\Gamma_{0\;0}^{0\;0}+\Gamma_{1\;2}^{1\;2}-\Gamma_{1\;0}^{1\;0}-\Gamma_{0\;2}^{0\;2}\right).
\lb{e64}
\eea
Differently from the $n=2$ case the new relations \rf{e63} and \rf{e64} are in general not consistent with relations \rf{e55} and \rf{e56}. This will restrict the integrability of the Hamiltonian \rf{e38} on special manifolds of the coupling constants $\Gamma_{m\;n}^{k\;l}$.

	As in the $n=2$ case, the use of \rf{e58}, with $n=3$, imply that an arbitrary amplitude of $|\Psi_{3,P}\rangle$ is proportional to $\mbox{Tr} \left[A_{k_l}A_{k_j}A_{k_m}E^{L}\Omega_P \right]$. The cyclic property of the trace supplemented  by the algebraic relations \rf{e42}, \rf{e47p} and \rf{e57} give us the relation that fix, the up to now free, complex spectral parameters $\{k_1,k_2,k_3 \}$, i. e.,
\beq
e^{ik_jL}=-\prod_{l=1}^n S(k_j,k_l), \;\;\; (j=1,\ldots,n),
\lb{e65}
\eeq
with $n=3$ and $S(k_j,k_l)$ given by \rf{e57}.

{\it {\bf  n = 4.}}
The relations coming from the amplitudes with collisions with up to three particles are solved by \rf{e47} and \rf{e51}, with energy and momentum given by \rf{e48}. The algebraic relations obtained are generalizations of \rf{e55}, \rf{e56}, \rf{e62} and \rf{e63}. A new relation relating the product of four matrices comes from the amplitude $f(x,x,x+1,x+1)$ where four particles are at colliding positions: 
\bea
&&\varepsilon_4\mbox{Tr} \left[E^{x-1}BE^{-1}BBE^{-1}BE^{L-x-1}\Omega_P \right]= \nonumber \\
&&-\Gamma_{0\;2}^{1\;1}\mbox{Tr} \left[E^{x-2}AABE^{-1}BE^{L-x-1}\Omega_P \right]-\Gamma_{2\;0}^{1\;1}\mbox{Tr} \left[E^{x-1}BE^{-1}BAAE^{L-x-2}\Omega_P \right] \nonumber \\
&&-\Gamma_{0\;2}^{2\;0}\mbox{Tr} \left[E^{x-2}BE^{-1}BEBE^{-1}BE^{L-x-1}\Omega_P \right]-\Gamma_{2\;0}^{0\;2}\mbox{Tr} \left[E^{x-1}BE^{-1}BEBE^{-1}BE^{L-x-2}\Omega_P \right] \nonumber \\
&&+\left(3\Gamma_{0\;0}^{0\;0}-\Gamma_{0\;2}^{0\;2}-\Gamma_{2\;0}^{2\;0}-\Gamma_{2\;2}^{2\;2} \right)\mbox{Tr} \left[E^{x-1}BE^{-1}BBE^{-1}BE^{L-x-1}\Omega_P \right]
\lb{e66}
\eea
Inserting \rf{e47}, \rf{e48} and \rf{e51} and using \rf{e55} and \rf{e56} in this last expression we obtain the algebraic relation with four spectral parameter matrices 
\bea
&&\sum_{q,r,s,t=1}^4 \left(D_4(k_q,k_r,k_s,k_t)B_{k_q}B_{k_r}B_{k_s}B_{k_t} \right. \nonumber \\
&&\left. +\Gamma_{2\;0}^{1\;1}e^{i(k_r+k_s+k_t)}A_{k_q}A_{k_r}B_{k_s}B_{k_t}+\Gamma_{0\;2}^{1\;1}e^{ik_t}B_{k_q}B_{k_r}A_{k_s}A_{k_t} \right)=0
\lb{e67}
\eea
where
\beq
D_4(k_1,k_2,k_3,k_4)=\Gamma_{0\;2}^{2\;0}+\Gamma_{2\;0}^{0\;2}e^{i(k_1+k_2+k_3+k_4)}-\left(\Gamma_{2\;2}^{2\;2}+\Gamma_{0\;0}^{0\;0}-\Gamma_{0\;2}^{0\;2}-\Gamma_{2\;0}^{2\;0} \right)e^{i(k_3+k_4)}.
\lb{e68}
\eeq
This relation will impose, in addition to \rf{e62} and \rf{e63}, further restrictions for the coupling constants $\Gamma_{m\;n}^{k\;l}$ of the general Hamiltonian \rf{e38}. The spectral parameters ($k_1,\ldots,k_4$), as in the previous cases, are fixed by the cyclic property of the trace, and are given by \rf{e65} with $n=4$.

{\it {\bf General n $>$ 4.}}
All the amplitudes produce relations that are solved by the spectral parameter matrices introduced in \rf{e47} and \rf{e51}. These matrices $\{A_{k_j} \}$ and $\{B_{k_j} \}$ ($j=1,\ldots,n$) should obey the set of algebraic relations
\beq
\left(A_{k_j} \right)^2=\left(B_{k_j} \right)^2=0 \;\;\; (j=1,\ldots,n)
\lb{e69}
\eeq
\beq
\sum_{p^{(2)}}N(k_{p_{1}},k_{p_{2}})A_{k_{p_{1}}}A_{k_{p_{2}}}=\sum_{p^{2}}C_1(k_{p_{1}},k_{p_{2}})B_{k_{p_{1}}}B_{k_{p_{2}}} \;\;\; (j \neq l=1,\ldots,n)
\lb{e70}
\eeq
\beq
\sum_{p^{(2)}}C_0(k_{p_{1}},k_{p_{2}})B_{k_{p_{1}}}B_{k_{p_{2}}}=\sum_{p_{2}}C_2(k_{p_{1}},k_{p_{2}})e^{ik_{p_{1}}}A_{k_{p_{1}}}A_{k_{p_{2}}} \;\;\; (j \neq l=1,\ldots,n)
\lb{e71}
\eeq
\bea
&&\sum_{p^{(3)}} \left(D_1(k_{p_{1}},k_{p_{2}},k_{p_{3}})B_{k_{p_{1}}}B_{k_{p_{2}}}A_{k_{p_{3}}}+\Gamma_{2\;0}^{1\;1}e^{i(k_{p_{2}}+k_{p_{3}})}A_{k_{p_{1}}}A_{k_{p_{2}}}A_{k_{p_{3}}}\right. \nonumber \\
&& \left. -\Gamma_{2\;1}^{1\;2}e^{i(k_{p_{2}}+k_{p_{3}})}A_{k_{p_{1}}}B_{k_{p_{2}}}B_{k_{p_{3}}} \right)=0,
\lb{e72}
\eea
\beq
\sum_{p^{(3)}} \left(D_2(k_{p_{1}},k_{p_{2}},k_{p_{3}})A_{k_{p_{1}}}B_{k_{p_{2}}}B_{k_{p_{3}}}+\Gamma_{0\;2}^{1\;1}e^{ik_{p_{3}}}A_{k_{p_{1}}}A_{k_{p_{2}}}A_{k_{p_{3}}}-\Gamma_{2\;1}^{1\;2}e^{ik_{p_{3}}}A_{k_{p_{1}}}B_{k_{p_{2}}}B_{k_{p_{3}}} \right)=0,
\lb{e73}
\eeq
\bea
&&\sum_{p^{(4)}} \left(D_4(k_{p_{1}},k_{p_{2}},k_{p_{3}},k_{p_{4}})B_{k_{p_{1}}}B_{k_{p_{2}} }B_{k_{p_{3}} }B_{k_{p_{4}} }+\Gamma_{2\;0}^{1\;1}e^{i(k_{p_{2}} +k_{p_{3}} +k_{p_{4}} )}A_{k_{p_{1}} }A_{k_{p_{2}} }B_{k_{p_{3}} }B_{k_{p_{4}} } \right. \nonumber \\
&&\left. +\Gamma_{0\;2}^{1\;1}e^{ik_{p_{4}} }B_{k_{p_{1}} }B_{k_{p_{2}} }A_{k_{p_{3}} }A_{k_{p_{4}} } \right)=0, 
\lb{e74}
\eea
where in the above expressions the sums are over the permutations $p^{(m)}$ ($m=2,3,4$) $\{ p_{1},\ldots,p_{m} \}$ of $m$ distinct integers ($j_1,\ldots,j_m$) taken from the set ($1,2,\ldots,n$). Since The general Hamiltonian \rf{e38} has only nearest-neighbor interactions we have no new relations. In fact only \rf{e72}-\rf{e74} produce constraints for the integrability of \rf{e38}. Exploring the fact that $C_1(k,k')$ is symmetric through the interchange $k \leftrightarrow k'$ we can use \rf{e70} to eliminate the matrices $\{B_{k} \}$ in \rf{e72}-\rf{e74}:
\bea
&&\sum_{p^{(3)}} \left\{D_1(k_{p_1},k_{p_2},k_{p_3})N(k_{p_1},k_{p_2})C_1(k_{p_2},k_{p_3})+e^{i(k_{p_2}+k_{p_3})}\left[\Gamma_{2\;0}^{1\;1}C_1(k_{p_2},k_{p_3})-\Gamma_{2\;1}^{1\;2}N(k_{p_2},k_{p_3}) \right] \right. \nonumber \\
&&\left. 
\times C_1(k_{p_1},k_{p_2}) \right\} 
 C_1(k_{p_1},k_{p_3})A_{k_{p_1}}A_{k_{p_2}}A_{k_{p_3}}=0
\lb{e75}
\eea
\bea
&&\sum_{p^{(3)}} \left\{D_2(k_{p_1},k_{p_2},k_{p_3})N(k_{p_2},k_{p_3})C_1(k_{p_1},k_{p_2})+e^{ik_{p_3}}\left[\Gamma_{0\;2}^{1\;1}C_1(k_{p_1},k_{p_2})-\Gamma_{1\;2}^{2\;1}N(k_{p_1},k_{p_2}) \right] \right\} \nonumber \\
&&\times C_1(k_{p_2},k_{p_2})C_1(k_{p_1},k_{p_3})A_{k_{p_1}}A_{k_{p_2}}A_{k_{p_3}}=0
\lb{e76}
\eea
\bea
&&\sum_{p^{(4)}} \left\{ D_4(k_{p_1},k_{p_2},k_{p_3},k_{p_4})N(k_{p_1},k_{p_2})N(k_{p_3},k_{p_4})+\Gamma_{2\;0}^{1\;1}e^{i(k_{p_2}+k_{p_3}+k_{p_4})}C_1(k_{p_1},k_{p_2})N(k_{p_3},k_{p_4}) \right. \nonumber \\
&&\left. +\Gamma_{0\;2}^{1\;1}e^{ik_{p_4}}N(k_{p_1},k_{p_2})C_1(k_{p_3},k_{p_4}) \right\}C_1(k_{p_1},k_{p_3})C_1(k_{p_1},k_{p_4})C_1(k_{p_2},k_{p_3})C_1(k_{p_2},k_{p_4}) \nonumber \\
&&\times A_{k_{p_{1}} }A_{k_{p_{2}} }A_{k_{p_{3}} }A_{k_{p_{4}} }=0.
\lb{e77}
\eea
The above relations should be consistent with the commutation relations
\beq
A_{k_j}A_{k_l}=S(k_j,k_l)A_{k_l}A_{k_j} \;\;\; (j \neq l), \;\;\; \left(A_{k_j} \right)^2=0 \;\;\; (j,l=1,2,\ldots,n).
\lb{e78}
\eeq
for any values of $k_j\in C$ ($j=1,\ldots,4$). The use of \rf{e78} in \rf{e75}-\rf{e77} enable us to express the left side as a polynomial on the variables $e^{ik_{p_j}}$ ($j=1,\ldots,4$). A sufficient condition for the integrability is obtained by requiring that all coefficients of this polynomial are zero.

        Although we did not consider the problem of finding all the possible solutions of \rf{e75}-\rf{e77} with \rf{e78} we verified that all the known exact integrable spin-$1$ chain with a single conservation law are solutions of these equations, defining properly an associative algebra. These are the cases of the following models. The  Fateev-Zamolodchikov model \cite{fateev} where
\bea
&&\Gamma_{0\;0}^{0\;0}=\Gamma_{2\;2}^{2\;2}=0, \;\;\; \Gamma_{0\;1}^{0\;1}=\Gamma_{2\;1}^{2\;1}=\Gamma_{0\;2}^{2\;0}=\Gamma_{2\;0}^{0\;2}=-1, \;\;\; \Gamma_{0\;1}^{1\;0}=\Gamma_{1\;0}^{0\;1}=\Gamma_{1\;2}^{2\;1}=\Gamma_{2\;1}^{1\;2}=1, \nonumber \\
&&\Gamma_{0\;2}^{1\;1}=\Gamma_{1\;1}^{0\;2}=\Gamma_{1\;1}^{2\;0}=\Gamma_{2\;0}^{1\;1}=2\cos(\gamma), \;\;\; \Gamma_{0\;2}^{0\;2}=\Gamma_{2\;0}^{2\;0}=-3+4\sin^{2}(\gamma), \nonumber \\
&&\Gamma_{1\;0}^{1\;0}=\Gamma_{1\;2}^{1\;2}=-1+4\sin^2(\gamma), \;\;\; \Gamma_{1\;1}^{1\;1}=-2+4\sin^{2}(\gamma),
\lb{e79}
\eea
and $\gamma$ is a free parameter; the Izergin-Korepin model \cite{izergin} where
\bea
&&\Gamma_{0\;1}^{1\;0}=\Gamma_{1\;0}^{0\;1}=\Gamma_{2\;1}^{1\;2}=\Gamma_{1\;2}^{2\;1}=1, \;\;\; \Gamma_{0\;0}^{0\;0}=\Gamma_{0\;1}^{0\;1}=\Gamma_{1\;0}^{1\;0}=0, \;\;\; \Gamma_{0\;2}^{2\;0}=\Gamma_{2\;0}^{0\;2}=\frac{\cosh(\gamma)}{\cosh(3\gamma)}, \nonumber \\
&&\Gamma_{1\;1}^{2\;0}=\Gamma_{2\;0}^{1\;1}=\frac{\cosh(2\gamma)}{\cosh(3\gamma)}e^{2\gamma}, \;\;\; \Gamma_{0\;2}^{1\;1}=\Gamma_{1\;1}^{0\;2}=-\frac{\cosh(2\gamma)}{\cosh(3\gamma)}e^{-2\gamma}, \;\;\; \Gamma_{1\;1}^{1\;1}=2\frac{\cosh(\gamma)\cosh(2\gamma)}{\cosh(3\gamma)}, \nonumber \\
&&\Gamma_{0\;2}^{0\;2}=\Gamma_{2\;0}^{2\;0}=\Gamma_{0\;2}^{2\;0}+2\Gamma_{1\;1}^{1\;1}\sinh^2(\gamma), \;\;\; \Gamma_{1\;2}^{1\;2}=\Gamma_{0\;2}^{0\;2}+\Gamma_{0\;2}^{2\;0}e^{-4\gamma}, \;\;\; \Gamma_{2\;1}^{2\;1}=\Gamma_{0\;2}^{0\;2}+\Gamma_{0\;2}^{2\;0}e^{4\gamma}, \nonumber \\
&&\Gamma_{2\;2}^{2\;2}=2\left(\Gamma_{0\;2}^{0\;2}+\cosh(4\gamma)\Gamma_{0\;2}^{2\;0} \right),
\lb{e80}
\eea
and $\gamma$ is also a free parameter; and the spin-$1$ Hamiltonian introduced in \cite{alcbar1} where
\bea
&&\Gamma_{0\;0}^{0\;0}=\Gamma_{0\;1}^{0\;1}=\Gamma_{1\;0}^{1\;0}=0, \;\;\; \Gamma_{0\;1}^{1\;0}=\Gamma_{1\;0}^{0\;1}=-1, \;\;\; \Gamma_{0\;2}^{2\;0}=\Gamma_{2\;0}^{0\;2}=-t, \;\;\; \Gamma_{1\;2}^{2\;1}=\Gamma_{2\;1}^{1\;2}=-\varepsilon, \nonumber \\
&&\Gamma_{1\;1}^{1\;1}=\varepsilon t, \;\;\; \Gamma_{2\;2}^{2\;2}=-\frac{2-\varepsilon}{t}, \;\;\; \Gamma_{1\;1}^{2\;0}=\Gamma_{2\;0}^{1\;1}=-e^{i\frac{\pi}{3}}\sqrt{t^2-1}, \;\;\; \Gamma_{0\;2}^{1\;1}=\Gamma_{1\;1}^{0\;2}=\varepsilon e^{-i\frac{\pi}{3}} \sqrt{t^2-1}, \nonumber \\
&&\Gamma_{1\;2}^{1\;2}=-\frac{1}{2}\left(\frac{2-\varepsilon}{t}-i\varepsilon \sqrt{3} t \right), \;\; \Gamma_{2\;1}^{2\;1}=-\frac{1}{2}\left(\frac{2-\varepsilon}{t}+i\varepsilon \sqrt{3} t \right), \nonumber \\
&&\Gamma_{0\;2}^{0\;2}=\Gamma_{2\;0}^{2\;0}=-\frac{1}{2}\left(\frac{2-\varepsilon}{t}+\varepsilon t \right),
\lb{e81}
\eea
with $\varepsilon=\pm 1$ and $t$ a free parameter.
In all these cases the eigenenergies and momentum are givem by \rf{e48} with $\{k_j \}$ given by \rf{e65} with the appropriate $S(k_j,k_l)$ matrix given in \rf{e57}, in agreement with the Bethe-ansatz solutions of these models. In order to conclude this section we mention that analogously to the generalyzed XXZ presented in section $2$, we can also produce straightforwardly a generalization of the models \rf{e79}-\rf{e81} where the spin $1$ and spin $0$ particles ($s^z$-basis) have now an effective size $s$ (1,2,\ldots) distinctly from the size $1$ of the particles with spin $-1$. A choice of the matrices as in \rf{e26}, i. e.,
\beq
A^{(s)}=\sum_{j=1}^nA_{k_j}^{(s)}E^{2-s}, \;\;\; B^{(s)}=\sum_{j=1}^nB_{k_j}^{(s)}E^{2-s},
\lb{e82}
\eeq
will give the same relations derived previously except that the spectral parameter, instead of been fixed by \rf{e65} is now given by
\beq
e^{ik_jL}=\prod_{l=1,\;l \neq j}^n\left(\frac{e^{ik_j}}{e^{ik_l}} \right)^{s-1}S(k_j,k_l), \;\;\; (j=1,\ldots,n).
\lb{e83}
\eeq


\section{ Models of spin $1$ with two conservation laws.}

        In this section we are going to formulate our matrix product ansatz for models describing the dynamics of two types of particles on the lattice, where the total number of particles of each type is conserved separately. Integrable models on this category are the spin-$1$ quantum chains like the anisotropic Sutherland model \cite{sutherland} or Perk-Schultz model \cite{perkshultz} and the t-J model \cite{schlo}, and the stochastic Hamiltonian that merge from the problem of asymmetric diffusion of particles hierarchically ordered \cite{alcbar4}. 
        Similarly as we did in section $1$ in order to illustrate the versatility of our matrix product  ansatz we are going to derive the extensions of the above models to the case where the two types of particles (type $1$ and $2$), or $s_z=1$ and $s_z=0$ in the spin representation of the models have arbitrary hard-core interactions. Particles of species $1$ ($2$) will have an effective size $s_1$ ($s_2$) that exclude the presence of particles on its rightmost $s_1-1$ sites ($s_2-1$ sites), where $s_1,s_2=1,2,\ldots.$. The above mentioned integrable models correspond to the particular case where $s_1=s_2=1$.

        Let us attach to each site $i=1,2,\ldots,L$ of the lattice, a site variable $Q_i$. that takes the values $Q_i=0$ if the site is empty or the values $Q_i=1$, $Q_i=2$ if the site is occupied by a particle of type $1$ (size $s_1$) and $2$ (size $s_2$) respectively. The general Hamiltonian we consider  has an  U($1$)$\otimes$U($1$) symmetry and governs the fluctuations of the configurations $\{Q_1,\ldots,Q_L \}$ on a ring of perimeter $L$:
\bea
&&H_{s_1,s_2}^{U(1)\otimes U(1)}=-\sum_{j=1}^LH_j +L\Gamma_{0\;0}^{0\;0} \nonumber \\
&&H_j={\cal{P}} \left[ \sum_{\alpha=1}^2 \left( \Gamma_{0\;\alpha}^{\alpha\;0} E_j^{0,\alpha}E_{j+1}^{\alpha,0}+\Gamma_{\alpha\;0}^{0\;\alpha} E_j^{\alpha,0}E_{j+1}^{0,\alpha} \right) \right. \nonumber \\
&&\left. +\sum_{\alpha=1}^2 \sum_{\beta=1}^2 \Gamma_{\beta\;\alpha}^{\alpha\;\beta} E_j^{\beta,\alpha}E_{j+s_{\beta}}^{\alpha,0}E_{j+s_{\alpha}}^{0,\beta}+ \sum_{\alpha=0}^2 \sum_{\beta=0}^2 \Gamma_{\alpha\;\beta}^{\alpha\;\beta} E_j^{\alpha,\alpha}E_{j+s_{\alpha}}^{\beta,\beta} \right] {\cal{P}},
\lb{e84}
\eea
where $E^{\alpha,\beta}$ ($\alpha,\beta=0,1,2$) are the usual $3\times 3$ Weyl matrix with $i,j$ elements $\left(E^{l,m} \right)_{i,j}=\delta_{l,i}\delta_{m,j}$ and $\Gamma_{n\;o}^{l\;m}$ are the coupling constants. The projector ${\cal{P}}$ in \rf{e84} projects out from the space of configurations those where the particles occupy forbidden positions due to their sizes. The last sum in \rf{e84} accounts for the ``static'' interactions while the first and second sums are the ``kinetic'' terms representing the motion and interchange of particles, respectively. The U($1$)$\otimes$U($1$) symmetry supplemented by the periodic boundary condition of \rf{e84} imply that the total number of particles $n_1,n_2=0,1,2,\ldots$ on class $1$ and $2$ as well the momentum $P=\frac{2\pi l}{L}$ ($l=0,1,\ldots$,L-1) are good quantum numbers.

        We want to formulate a matrix product  ansatz for the eigenvectors $\Psi_{n_1,n_2,P}$ of the eigenvalue equation
\beq
H_{s_1,s_2}^{U(1)\otimes U(1)}|\Psi_{n_1,n_2,P}\rangle=\varepsilon_{n_1,n_2}|\Psi_{n_1,n_2,P}\rangle
\lb{e85}
\eeq
belonging to the eigensector labeled by ($n_1,n_2,P$). These eigenvectors are given by
\beq
|\Psi_{n_1,n_2,P}\rangle=\sum_{\{Q \}}\sum_{\{x \}} f(x_1,Q_1;\ldots;x_n,Q_n)|x_1,Q_1;\ldots;x_n,Q_n\rangle,
\lb{e86}
\eeq
where the kets $|x_1,Q_1;\ldots;s_n,Q_n\rangle$ denote the configurations with particles of type $Q_i$ ($Q_i=1,2$) located at the positions $x_i$ ($x_i=1,\ldots,L$). The total number of particles is $n=n_1+n_2$. The summation $\{Q \}=\{Q_1,\ldots,Q_n\}$ extends over all the permutations of $n$ numbers $\{1,2 \}$ in which $n_1$ terms have value $1$ and $n_2$ terms the value $2$, while the summation $\{x \}=\{x_1,\ldots,x_n \}$ extends, for each permutation $\{Q \}$, into the set of the nondecreasing integers satisfying
\bea
&&x_{i+1}\ge x_i+s_{Q_i}=1,\ldots,n-1 \nonumber \\
&&s_{Q_1} \le x_n-x_1\le L-s_{Q_n}.
\lb{87}
\eea

	The matrix product  ansatz we propose asserts that the amplitudes of an arbitrary eigenfunction \rf{e86} are given in terms of traces of the matrix product
\beq
f(x_1,Q_1;\ldots;x_n,Q_n)=\mbox{Tr} \left[E^{x_1-1}Y^{(Q_1)}E^{x_2-x_1-1}Y^{(Q_2)}\cdots E^{x_n-x_{n-1}-1}Y^{(Q_n)}E^{L-x_n}\Omega_P \right].
\lb{e88}
\eeq
The matrices $Y^{(Q)}$ are associated to the particles of type $Q$ (Q=1,2). As in the previous sections the matrix $E$ is associated to the vacant sites and the matrix $\Omega_P$ satisfying
\beq
E\Omega_P=e^{-iP}\Omega_PE, \;\;\; Y^{(Q)}\Omega_P=e^{-iP}\Omega_PY^{(Q)} \;\;\; (Q=1,2)
\lb{e89}
\eeq
ensures (see \rf{e6}) the momentum $P=\frac{2\pi}{L}l$ ($l=0,1,\ldots,L-1$) of the eigenvector. Let us consider initially the simpler cases where $n=1$ and $n=2$.

{\it {\bf  n = 1.}}
We have distinct equations depending on the type $Q=1,2$ of the particle. The eigenvalue equation \rf{e85} give us
\bea
&&\varepsilon^{(Q)} \mbox{Tr} \left[E^{x-1}Y^{(Q)}E^{L-x}\Omega_P \right]=-\Gamma_{0\;Q}^{Q\;0}\mbox{Tr} \left[E^{x-2}Y^{(Q)}E^{L-x+1}\Omega_P \right]-\Gamma_{Q\;0}^{0\;Q}\mbox{Tr} \left[E^{x}Y^{(Q)}E^{L-x-1}\Omega_P \right] \nonumber \\
&&+ \left(2\Gamma_{0\;0}^{0\;0}-\Gamma_{0\;Q}^{0\;Q}-\Gamma_{Q\;0}^{Q\;0} \right)\mbox{Tr} \left[E^{x-1}Y^{(Q)}E^{L-x}\Omega_P \right], 
\lb{e90}
\eea
where $\varepsilon^{(1)}=\varepsilon_{1,0}$, $\varepsilon^{(2)}=\varepsilon_{0,1}$ are the eigenvalues. As in the previous sections a convenient solution is obtained by introducing the spectral parameter dependent matrices
\beq
Y^{(Q)}=Y_k^{(Q)}E^{2-s_Q} \;\;\; (Q=1,2),
\lb{e91}
\eeq
with $k \in C$, that satisfy the commutation relation with the matrix $E$
\beq
EY_k^{(Q)}=e^{ik}Y_k^{(Q)}E \;\;\; (Q=1,2).
\lb{e92}
\eeq
Inserting \rf{e91} and \rf{e92} into \rf{e90} we obtain
\bea
&&\varepsilon^{(Q)}(k)=-\left(\Gamma_{0\;Q}^{Q\;0}e^{-ik}+\Gamma_{Q\;0}^{0\;Q}e^{ik}-2\Gamma_{0\;0}^{0\;0}+\Gamma_{0\;Q}^{0\;Q}+\Gamma_{Q\;0}^{Q\;0} \right) \;\;\; (Q=1,2) \nonumber \\
&&P=k=\frac{2\pi}{L}l \;\;\; (l=0,1,\ldots,L-1).
\lb{e93}
\eea

{\it {\bf  n = 2.}}
For two particles of types $Q_1$ and $Q_2$ ($Q_1,Q_2=1,2$) on the lattice we have two types of relations. The eigenvalue equation applied to the components where the particles of class $Q_1$ and $Q_2$ are at positions ($x_1$, $x_2$) with $x_2>x_1+s_{Q_1}$ give us the generalization of \rf{e90}
\bea
&&\varepsilon^{(Q_1,Q_2)} \mbox{Tr} \left[E^{x_1-1}Y^{(Q_1)}E^{x_2-x_1-1}Y^{(Q_2)}E^{L-x_2}\Omega_P \right]= \nonumber \\
&&-\Gamma_{0\;Q_1}^{Q_1\;0}\mbox{Tr} \left[E^{x_1-2}Y^{(Q_1)}E^{x_2-x_1}Y^{(Q_2)}E^{L-x_2}\Omega_P \right]-\Gamma_{Q_1\;0}^{0\;Q_1}\mbox{Tr} \left[E^{x_1}Y^{(Q_1)}E^{x_2-x_1-2}Y^{(Q_2)}E^{L-x_2}\Omega_P \right] \nonumber \\
&&-\Gamma_{0\;Q_2}^{Q_2\;0}\mbox{Tr} \left[E^{x_1-1}Y^{(Q_1)}E^{x_2-x_1-2}Y^{(Q_2)}E^{L-x_2+1}\Omega_P \right] \nonumber \\
&&-\Gamma_{Q_2\;0}^{0\;Q_2}\mbox{Tr} \left[E^{x_1-1}Y^{(Q_1)}E^{x_2-x_1}Y^{(Q_2)}E^{L-x_2-1}\Omega_P \right] \nonumber \\
&&+ \left(4\Gamma_{0\;0}^{0\;0}-\Gamma_{0\;Q_1}^{0\;Q_1}-\Gamma_{Q_1\;0}^{Q_1\;0} -\Gamma_{0\;Q_2}^{0\;Q_2}-\Gamma_{Q_2\;0}^{Q_2\;0}\right)\mbox{Tr} \left[E^{x_1-1}Y^{(Q_1)}E^{x_2-x_1-1}Y^{(Q_2)}E^{L-x_2}\Omega_P \right],
\lb{e94}
\eea
while the components where the particles are at the colliding positions ($x_1=x$, $x_2=x_1+s_{Q_1}$) give us
\bea
&&\varepsilon^{(Q_1,Q_2)} \mbox{Tr} \left[E^{x-1}Y^{(Q_1)}E^{s_{Q_1}-1}Y^{(Q_2)}E^{L-x-s_{Q_1}}\Omega_P \right]= \nonumber \\
&&-\Gamma_{0\;Q_1}^{Q_1\;0}\mbox{Tr} \left[E^{x-2}Y^{(Q_1)}E^{s_{Q_1}}Y^{(Q_2)}E^{L-x-s_{Q_1}}\Omega_P \right] \nonumber \\
&&-\Gamma_{Q_2\;0}^{0\;Q_2}\mbox{Tr} \left[E^{x-1}Y^{(Q_1)}E^{s_{Q_1}}Y^{(Q_2)}E^{L-x-s_{Q_1}-1}\Omega_P \right] \nonumber \\
&& -\Gamma_{Q_1\;Q_2}^{Q_2\;Q_1}\mbox{Tr} \left[E^{x-1}Y^{(Q_2)}E^{s_{Q_2}-1}Y^{(Q_1)}E^{L-x-s_{Q_2}}\Omega_P \right] \nonumber \\
&&+ \left(3\Gamma_{0\;0}^{0\;0}-\Gamma_{0\;Q_1}^{0\;Q_1}-\Gamma_{Q_2\;0}^{Q_2\;0} -\Gamma_{Q_1\;Q_2}^{Q_1\;Q_2}\right)\mbox{Tr} \left[E^{x-1}Y^{(Q_1)}E^{s_{Q_1}-1}Y^{(Q_2)}E^{L-x-s_{Q_1}}\Omega_P \right].
\lb{e95}
\eea

	Let us consider initially the case where the particles are of same type. In this case we have exactly the same situation as in the XXZ chain considered in section $1$, and a solution of \rf{e94}-\rf{e95} is obtained by identifying $Y^{(Q)}$ as composed by two spectral parameter dependent matrices $Y_{k_1}^{(Q)}$ and $Y_{k_2}^{(Q)}$, i. e.,
\beq
Y^{(Q)}=\sum_{j=1}^nY_{k_j}^{(Q)}E^{2-s_Q} \;\;\; \mbox{with} \;\;\; EY_{k_j}^{(Q)}=e^{ik_j}Y_{k_j}^{(Q)}E, \;\;\; \left(Y_{k_j}^{(Q)} \right)^2=0 \;\;\; (Q=1,2), 
\lb{e96}
\eeq
with $n=2$. These last relations, when inserted in \rf{e94} give us the energy and momentum in terms of the spectral parameters $k_j$ ($j=1,2$)
\beq
\varepsilon^{(Q,Q)}=\sum_{j=1}^n\varepsilon^{(Q)}(k_j), \;\;\; P=\sum_{j=1}^nk_j,
\lb{e97}
\eeq
where $n=2$ and $\varepsilon^{(Q)}(k)$ is given by \rf{e93}. Using \rf{e96} and \rf{e97} in \rf{e95} we obtain the relations
\beq
Y_{k_j}^{(Q)}Y_{k_l}^{(Q)}=S_{Q\;Q}^{Q\;Q}(k_j,k_l)Y_{k_l}^{(Q)}Y_{k_j}^{(Q)} \;\;\; (j \ne l), \;\;\; \left(Y_{k_j}^{(Q)} \right)^2=0, \;\;\; (1 \le j,l \le n)
\lb{e98}
\eeq
where
\beq
S_{Q\;Q}^{Q\;Q}(k_j,k_l)=-\frac{\Gamma_{0\;Q}^{Q\;0}+\Gamma_{Q\;0}^{0\;Q}e^{i(k_j+k_l)}-\left(\Gamma_{0\;0}^{0\;0}+\Gamma_{Q\;Q}^{Q\;Q}-\Gamma_{Q\;0}^{Q\;0}-\Gamma_{0\;Q}^{0\;Q} \right)e^{ik_j}}{\Gamma_{0\;Q}^{Q\;0}+\Gamma_{Q\;0}^{0\;Q}e^{i(k_j+k_l)}-\left(\Gamma_{0\;0}^{0\;0}+\Gamma_{Q\;Q}^{Q\;Q}-\Gamma_{Q\;0}^{Q\;0}-\Gamma_{0\;Q}^{0\;Q} \right)e^{ik_l}} \;\;\; (Q=1,2).
\lb{e99}
\eeq

	Let us consider now the case where the particles are of distinct species. The distinguibility of the particles allows two type of solutions of \rf{e94}-\rf{e95}. We may tray a standard solution as in \rf{e96} (case a) where each of the matrices $Y^{(Q)}$ ($Q=1,2$) are composed by two spectral parameter matrices, with the same value of the spectral parameters $k_1,k_2$ (see \rf{e96}) or alternatively (case b) we may consider a special solution where each $Y^{(Q)}$ is composed by a single spectral parameter matrix, with a distinct spectral parameter, i. e.,
\beq
Y^{(1)}=Y_{k_1}^{(1)}E^{2-s_1}, \;\;\; Y^{(2)}=Y_{k_2}^{(2)}E^{2-s_2} \;\;\; \mbox{with} \;\;\; EY_{k_j}^{(j)}=e^{ik_j}Y_{k_j}^{(j)}E \;\;\; (j=1,2).
\lb{e100}
\eeq
In case b \rf{e94} give us the energy and momentum as in \rf{e97} while \rf{e95} give us two independent equations:
\bea
&&\left[ -\left(\Gamma_{0\;Q_2}^{Q_2\;0}+\Gamma_{Q_1\;0}^{0\;Q_1}e^{i(k_1+k_2)} \right)+\left(\Gamma_{0\;0}^{0\;0}+\Gamma_{Q_1\;Q_2}^{Q_1\;Q_2}-\Gamma_{Q_1\;0}^{Q_1\;0}-\Gamma_{0\;Q_2}^{0\;Q_2}\right)e^{ik_{Q_2}} \right]Y_{k_{Q_1}}^{(Q_1)}Y_{k_{Q_2}}^{(Q_2)} \nonumber \\
&&+\Gamma_{Q_1\;Q_2}^{Q_2\;Q_1}e^{ik_{Q_2}}Y_{k_{Q_2}}^{(Q_2)}Y_{k_{Q_1}}^{(Q_1)}=0 \;\;\; (Q_1 \ne Q_2=1,2).
\lb{e101}
\eea
Since at this level we want to keep $k_1$ and $k_2$ as free complex parameters \rf{e101} imply special choices of the coupling constants $\Gamma_{k\;l}^{m\;n}$ of \rf{e84}, for example\footnote{Another equivalent choices are obtained by the interchange $1\leftrightarrow 2$ in \rf{e102}-\rf{e104}.},
\beq
\Gamma_{0\;2}^{2\;0}=\Gamma_{1\;0}^{0\;1}=\Gamma_{1\;2}^{2\;1}=0, \;\;\; \Gamma_{0\;0}^{0\;0}+\Gamma_{1\;2}^{1\;2}=\Gamma_{1\;0}^{1\;0}+\Gamma_{0\;2}^{0\;2},
\lb{e102}
\eeq
that gives
\bea
&&Y_{k_1}^{(1)}Y_{k_2}^{(2)}=S_{1\;2}^{1\;2}(k_1,k_2)Y_{k_2}^{(2)}Y_{k_1}^{(1)}, \nonumber \\
&&Y_{k_2}^{(2)}Y_{k_1}^{(1)}=S_{2\;1}^{2\;1}(k_2,k_1)Y_{k_1}^{(1)}Y_{k_2}^{(2)}
\lb{e103}
\eea
where
\beq
S_{1\;2}^{1\;2}(k_1,k_2)=\frac{1}{S_{2\;1}^{2\;1}(k_2,k_1)}=\frac{\Gamma_{0\;1}^{1\;0}+\Gamma_{2\;0}^{0\;2}e^{i(k_1+k_2)}-\left(\Gamma_{0\;0}^{0\;0}+\Gamma_{2\;1}^{2\;1}-\Gamma_{2\;0}^{2\;0}-\Gamma_{0\;1}^{0\;1}\right)e^{ik_1}}{\Gamma_{2\;1}^{1\;2}e^{ik_1}}.
\lb{e104}
\eeq
Let us consider case a. A solution of \rf{e94} and \rf{e95} where $Y^{(Q)}$ ($Q=1,2$) is given in terms of two spectral parameter matrices $\{Y_{k_j}^{(Q)} \}$ as in \rf{e96} is possible only if these matrices satisfy
\beq
Y_{k_j}^{(Q_1)}Y_{k_j}^{(Q_2)}=0 \;\;\; j=1,2 \;\;\; (Q_1,Q_2=1,2),
\lb{e105}
\eeq
and the Hamiltonian \rf{e84} has its coupling constants restricted to
\beq
\Gamma_{0\;1}^{1\;0}=\Gamma_{0\;2}^{2\;0}, \;\;\; \Gamma_{1\;0}^{0\;1}=\Gamma_{2\;0}^{0\;2}.
\lb{e106}
\eeq
With \rf{e105} and \rf{e106} the energy and momentum are given by 
\beq
\varepsilon^{(Q_1,Q_2)}=\varepsilon^{(Q_1)}(k_1)+\varepsilon^{(Q_2)}(k_2), \;\;\; P=k_1+k_2,
\lb{e107}
\eeq
respectively. Inserting \rf{e96} and \rf{e107} into \rf{e95} we obtain algebraic relations that can be written in a matrix form
\beq
\label{e108}
\sum_{l,m=1}^2
$$\left[ \matrix{  {\cal D}_{l,m} + v_{Q_2,Q_1} e^{ik_m} &
\Gamma_{Q_1\;Q_2}^{Q_2\;Q_1}e^{ik_m}  \cr
  \Gamma_{Q_2\;Q_1}^{Q_1\;Q_2} e^{ik_m} & {\cal
D}_{l,m} + v_{Q_1,Q_2} e^{ik_m}   }       \right]
\left[ \matrix{ Y_{k_l}^{(Q_1)}Y_{k_m}^{(Q_2)} \cr
  Y_{k_l}^{(Q_2)}Y_{k_m}^{(Q_1)}     }
\right] = 0.$$
\eeq
where 
\beq
{\cal D}_{l,m}=-\left(\Gamma_{0\;Q_1}^{Q_1\;0}+\Gamma_{Q_1\;0}^{0\;Q_1}e^{i(k_l+k_m)} \right), \;\;\; v_{Q_2,Q_1}=\Gamma_{0\;0}^{0\;0}+\Gamma_{Q_1\;Q_2}^{Q_1\;Q_2}-\Gamma_{Q_1\;0}^{Q_1\;0}-\Gamma_{0\;Q_2}^{0\;Q_2} \;\;\; (Q_1, Q_2=1,2).
\lb{e109}
\eeq
This last relation can be rearranged straightforwardly (see e.g. \cite{alcbar4}) giving us ($Q_1 \ne Q_2$)
\beq
Y_{k_l}^{(Q_1)}Y_{k_m}^{(Q_2)}=S_{Q_1\;Q_2}^{Q_1\;Q_2}(k_l,k_m)Y_{k_m}^{(Q_2)}Y_{k_l}^{(Q_1)}+S_{Q_2\;Q_1}^{Q_1\;Q_2}(k_l,k_m)Y_{k_m}^{(Q_1)}Y_{k_l}^{(Q_2)}
\lb{e110}
\eeq
where
\bea
&&S_{Q_2\;Q_1}^{Q_1\;Q_2}(k_l,k_m)=-\left\{ 1-\frac{e^{ik_l}-e^{k_m}}{\Delta}\left[\left({\cal D}_{l,m} + v_{Q_1,Q_2} e^{ik_m} \right) v_{Q_2,Q_1}-\Gamma_{Q_1\;Q_2}^{Q_2\;Q_1}\Gamma_{Q_2\;Q_1}^{Q_1\;Q_2}e^{ik_m} \right] \right\} \nonumber \\
&&S_{Q_1\;Q_2}^{Q_1\;Q_2}(k_l,k_m)=-\left\{ 1-\frac{e^{ik_l}-e^{k_m}}{\Delta}\left[\left({\cal D}_{l,m} + v_{Q_1,Q_2} e^{ik_m} \right)\Gamma_{Q_1\;Q_2}^{Q_2\;Q_1}-\Gamma_{Q_1\;Q_2}^{Q_2\;Q_1} v_{Q_1,Q_2}e^{ik_m} \right] \right\}
\lb{e111}
\eea
and
\beq
\Delta=\left( {\cal D}_{l,m} + v_{Q_2,Q_1} e^{ik_m} \right) \left({\cal D}_{l,m} + v_{Q_1,Q_2} e^{ik_m} \right)-\Gamma_{Q_1\;Q_2}^{Q_2\;Q_1}\Gamma_{Q_2\;Q_1}^{Q_1\;Q_2}e^{i2k_m}.
\lb{e112}
\eeq
In all cases, for $Q_1=Q_2$ (see \rf{e98} and \rf{e99}) or $Q_1 \ne Q_2$ in case b (see \rf{e102}-\rf{e104}) or case a (see \rf{e110}-\rf{e112}) the cyclic property of the traces in our matrix product  ansatz \rf{e88} will fix the spectral parameters. Instead of producing these equations for $n=2$ let us consider the case of general values of $n$.

{\it {\bf  General n.}}
We now consider the case of arbitrary numbers $n_1$, $n_2$ of particles of type $1$ and $2$ ($n=n_1+n_2$). The eigenvalue equation \rf{e85} when applied to the components of the eigenfunction $|\Psi_{n_1,n_2,P}\rangle$ where all the particles are not at colliding positions, give us a generalization of \rf{e90} and \rf{e94} that can be solved in two distinct ways as in the case $n=2$. In case b, where the coupling constants satisfy \rf{e102}, we identify the matrices $Y^{(Q)}$ ($Q=1,2$) as composed by the spectral dependent matrices
\beq
Y^{(1)}=\sum_{j=1}^{n_1} Y_{k_j}^{(1)}E^{2-s_1}, \;\;\; Y^{(2)}=\sum_{j=n_1+1}^{n} Y_{k_j}^{(2)}E^{2-s_2},
\lb{e113}
\eeq
where $Y_{k_j}^{(Q)}$ satisfy
\beq
EY_{k_j}^{(Q)}=e^{ik_j}Y_{k_j}^{(Q)}E, \;\;\; \left(Y_{k_j}^{(Q)} \right)^2=0, \;\;\; n_{Q-1} < j \le n_Q+(Q-1)n_{Q-1} , \;\;\; (Q=1,2),
\lb{e114}
\eeq
with $n_0=0$. On the other hand in case a, where the coupling constants satisfy \rf{e106}, the matrices $Y^{(Q)}$ are given by
\beq
Y^{(Q)}=\sum_{j=1}^n Y_{k_j}^{(Q)}E^{2-s_Q}, \;\;\; \mbox{with} \;\;\; EY_{k_j}^{(Q)}=e^{ik_j}Y_{k_j}^{(Q)}E \;\;\; 1 \le j \le n.
\lb{e115}
\eeq
In cases a and b the energy and momentum, in terms of the spectral parameters $\{k_j \}$, are given by
\beq
\varepsilon_{n_1,n_2}=\sum_{j=1}^{n_1}\varepsilon^{Q_1}(k_j)+\sum_{j=n_1+1}^n\varepsilon^{Q_2}(k_j), \;\;\; P=\sum_{j=1}^nk_j.
\lb{e116}
\eeq

	The amplitudes of $|\Psi_{n_1,n_2,P}\rangle$ where a pair of particles of types $Q_1$ and $Q_2$ are located at the closest positions $x_i$ and $x_{i+1}=x_i+s_{Q_1}$ will give us the following algebraic relations. In case b we obtain
\bea
&&Y_{k_j}^{(Q_1)}Y_{k_l}^{(Q_2)}=S_{Q_1\;Q_2}^{Q_1\;Q_2}(k_j,k_l)Y_{k_l}^{(Q_2)}Y_{k_j}^{(Q_1)} \;\;\; (Q_1,Q_2=1,2) \;\;\; (k_j \ne k_l), \nonumber \\
&&n_{Q_1-1} < j \le n_{Q_1}+(Q_1-1)n_{Q_1-1}, \;\;\; n_{Q_2-1} < l \le n_{Q_2}+(Q_2-1)n_{Q_2-1}, \;\;\; n_0=0,
\lb{e117}
\eea
where  the algebraic structure constants are the diagonal $S$-matrix with non-zero elements given by 
\bea
&&S_{Q\;Q}^{Q\;Q}(k_j,k_l)=-\frac{\Gamma_{0\;Q}^{Q\;0}+\Gamma_{Q,0}^{0,Q}e^{i(k_j+k_l)}-\left(\Gamma_{0\;0}^{0\;0}+\Gamma_{Q\;Q}^{Q\;Q}-\Gamma_{Q\;0}^{Q\;0}-\Gamma_{0\;Q}^{0\;Q} \right)e^{ik_j}}{\Gamma_{0\;Q}^{Q\;0}+\Gamma_{Q,0}^{0,Q}e^{i(k_j+k_l)}-\left(\Gamma_{0\;0}^{0\;0}+\Gamma_{Q\;Q}^{Q\;Q}-\Gamma_{Q\;0}^{Q\;0}-\Gamma_{0\;Q}^{0\;Q} \right)e^{ik_l}},  \nonumber \\
&&S_{1\;2}^{1\;2}(k_j,k_l)=\frac{1}{S_{2\;1}^{2\;1}(k_l,k_j)}=\frac{\Gamma_{0\;1}^{1\;0}+\Gamma_{2\;0}^{0\;2}e^{i(k_j+k_l)}-\left(\Gamma_{0\;0}^{0\;0}+\Gamma_{2\;1}^{2\;1}-\Gamma_{2\;0}^{2\;0}-\Gamma_{0\;1}^{0\;1} \right)e^{ik_j}}{\Gamma_{2\;1}^{1\;2}e^{ik_j}}.
\lb{e118}
\eea
In case a, where the coupling constants satisfy \rf{e106} we obtain
\beq
Y_{k_l}^{(Q_1)}Y_{k_m}^{(Q_2)}=\sum_{Q_1'=1}^2\sum_{Q_2'=1}^2S_{Q_2'\;Q_1'}^{Q_1\;Q_2}(k_l,k_m)Y_{k_m}^{(Q_1')}Y_{k_l}^{(Q_2')} \;\;\; (k_l \ne k_m), \;\;\; 1 \le l,m \le n,
\lb{e119}
\eeq
where the non vanishing values of the non-diagonal $S$-matrix are the six values given in \rf{e111} and \rf{e99} with the choice \rf{e102}.

	For arbitrary amplitudes we have in our matrix product  ansatz \rf{e88} a product of $n$ matrices $\{Y_{k_j}^{(Q)} \}$. Our  ansatz will be valid only if the relations \rf{e117} in case b or \rf{e119} in case a provide a unique relation among these products, otherwise $|\Psi_{n_1,n_2,P}\rangle$ in not properly defined. This means, for example, that the products $\cdots Y_{k_1}^{(\alpha)}Y_{k_2}^{(\beta)}Y_{k_3}^{(\gamma)}\cdots$ and $\cdots Y_{k_3}^{(\gamma)}Y_{k_2}^{(\beta)}Y_{k_1}^{(\alpha)}\cdots$ should be uniquely related. Since we can relate then either by performing the commutations in the order $\alpha \beta \gamma \rightarrow \beta \alpha \gamma \rightarrow \beta \gamma \alpha \rightarrow \gamma \beta \alpha$ or $\alpha \beta \gamma \rightarrow \alpha \gamma \beta \rightarrow \gamma \alpha \beta \rightarrow \gamma \beta \alpha$ the structure constants $S_{\gamma\;\gamma'}^{\alpha\;\alpha'}$ of the algebraic relations \rf{e117} and \rf{e119} should satisfy
\bea
&&\sum_{\gamma,\gamma',\gamma''=1}^2 S_{\gamma\;\gamma'}^{\alpha\;\alpha'}(k_1,k_2)S_{\beta\;\gamma''}^{\gamma\;\alpha''}(k_1,k_3)S_{\beta'\;\beta''}^{\gamma'\;\gamma''}(k_2,k_3)= \nonumber \\
&&\sum_{\gamma,\gamma',\gamma''=1}^2 S_{\gamma'\;\gamma''}^{\alpha'\;\alpha''}(k_2,k_3)S_{\gamma\;\beta''}^{\alpha\;\gamma''}(k_1,k_3)S_{\beta\;\beta'}^{\gamma\;\gamma'}(k_1,k_2),
\lb{e120}
\eea
for $\alpha, \alpha', \alpha'', \beta, \beta', \beta''=1,2$. This last constraint is just the Yang-Baxter relation \cite{yang2,baxter} of the $S$-matrix defined in \rf{e117} and \rf{e119}. Actually the condition \rf{e120} is enough to ensure that any matrix product of spectral matrices $\{Y_{k_j}^{(Q)} \}$ is uniquely related and it implies the associativity of the algebra of the operators $\{Y_{k_j}^{(Q)} \}$.

	In case b, the $S$-matrix given by \rf{e118} is diagonal and the Yang-Baxter relation \rf{e120} is satisfied trivially, so that the only restriction in the coupling constants of the Hamiltonian \rf{e84} is given by \rf{e102}. In case a the $S$-matrix given in \rf{e99} and \rf{e111} is non diagonal and the Yang-Baxter relation \rf{e120} produces strong constraints in the allowed couplings of the Hamiltonian \rf{e84}.

	We did not consider in this paper the problem of finding all the possible coupling constants $\Gamma_{j\;l}^{m\;n}$ in  \rf{e84} that renders the algebra \rf{e119} associative, or equivalently, to find all the solutions of the Yang-Baxter relations  \rf{e120}. It is important to stress that our matrix product  ansatz \rf{e88} produce the same relations \rf{e119} independently of the hard-core sizes $s_1$ and $s_2$ of the particles of type $1$ and $2$, respectively. Solutions of \rf{e120} were presented in the literature along the years. The solution where 
\bea
&&\Gamma_{\beta\;\alpha}^{\alpha\;\beta}=1 \;\;\; (\alpha \ne \beta), \;\;\; \Gamma_{\alpha\;\alpha}^{\alpha\;\alpha}=\epsilon_{\alpha} \cosh(\gamma), \nonumber \\
&&\Gamma_{\alpha\;\beta}^{\alpha\;\beta}={\mbox{sign}}(\alpha-\beta)\sinh(\gamma) \;\;\; (\alpha \ne \beta),
\lb{e121}
\eea
where $\gamma$ is a free parameter and $\epsilon_1, \epsilon_2, \epsilon_3=\pm 1$, corresponds to the anisotropic Perk-Schultz model \cite{perkshultz}. The isotropic model obtained by setting $\gamma=0$ and $\epsilon_1, \epsilon_2, \epsilon_3=\pm 1$ is the SU($3$) Sutherland model \cite{sutherland}. The solution \rf{e121} with $\epsilon_1=-\epsilon_2=\epsilon_3=1$ give us the anisotropic supersymmetric t-J model\footnote{A Jordan-Wigner fermionization of the Hamiltonian \rf{e84} with $s_1=s_2=1$ and coupling constants \rf{e121} give us the anisotropic supersymmetric t-J model \cite{schlo}. }. A solution of \rf{e120} is also known \cite{bjp4} for the case where
\beq
\Gamma_{\beta\;\alpha}^{\alpha\;\beta}=-\Gamma_{\alpha\;\beta}^{\alpha\;\beta}=q^{{\mbox{sign}}(\alpha-\beta)} \;\;\; (\alpha \ne \beta), \;\;\; \Gamma_{\alpha\;\alpha}^{\alpha\;\alpha}=0 \;\;\; (\alpha, \beta=0,1,2),
\lb{e121p}
\eeq
with $q$ real and $q \ge 1$. The Hamiltonian \rf{e84} with the couplings \rf{e121p} describes the time fluctuations in the asymmetric diffusion problem of two species of particles hierarchically ordered (asymmetry $q$). A generalization of this problem, treated through the matrix product  ansatz is presented elsewhere \cite{bjp3}.

        In order to complete our solutions through the matrix product  ansatz \rf{e88} we should fix the spectral parameters ($k_1,\ldots, k_n$). Using the algebraic relations \rf{e114} or \rf{e115} an arbitrary amplitude is proportional to 
\beq
\mbox{Tr} \left[ Y_{k_1}^{(1)}\cdots  Y_{k_{n_1}}^{(1)}Y_{k_{n_1+1}}^{(2)}\cdots Y_{k_n}^{(2)} E^{L-n_1(s_1-1)-n_2(s_2-1)}\Omega_P \right].
\lb{e122}
\eeq
The cyclic property of the trace fix the spectral parameters.

        Let us consider case b, where the coupling constants satisfy \rf{e102}. The commutation relations \rf{e89}, \rf{e114} and \rf{e117} give us 
\bea
&&e^{ik_jL}=\left[\prod_{l=1}^{n_1}S_{1\;1}^{1\;1}(k_j,k_l)\left(\frac{e^{ik_j}}{e^{ik_l}} \right)^{s_1-1} \right]\prod_{l=n_1+1}^{n}S_{1\;2}^{1\;2}(k_j,k_l)\left(\frac{e^{ik_j}}{e^{ik_l}} \right)^{s_2-1} \;\;\; (1 \le j \le n_1) \nonumber \\
&&e^{ik_jL}=\left[\prod_{l=n_1+1}^{n}S_{2\;2}^{2\;2}(k_j,k_l)\left(\frac{e^{ik_j}}{e^{ik_l}} \right)^{s_2-1} \right]\prod_{l=1}^{n_1}S_{2\;1}^{2\;1}(k_j,k_l)\left(\frac{e^{ik_j}}{e^{ik_l}} \right)^{s_1-1} \;\;\; (n_1 < j \le n),
\lb{e123}
\eea
where the $S$-matrix is given by \rf{e118}. The energy and momentum are obtained by inserting the solutions of \rf{e123} into \rf{e116}. The Hamiltonian \rf{e84} that is exactly integrable in case b (see \rf{e102}) is given by 
\bea
H^{(b)}&=&-\sum_{j=1}^L {\cal{P}} \left\{ \Gamma_{0\;1}^{1\;0}E_j^{0\;1}E_{j+1}^{1\;0}+\Gamma_{2\;0}^{0\;2}E_j^{2\;0}E_{j+1}^{0\;2}+\Gamma_{2\;1}^{1\;2}E_j^{2\;1}E_{j+s_2}^{1\;0}E_{j+s_1}^{0\;2}+\Gamma_{1\;1}^{1\;1}E_j^{1\;1}E_{j+s_1}^{1\;1} \right. \nonumber \\
&&+\Gamma_{2\;2}^{2\;2}E_j^{2\;2}E_{j+s_2}^{2\;2}+\Gamma_{0\;1}^{0\;1}E_j^{0\;0}E_{j+1}^{1\;1}+\Gamma_{1\;0}^{1\;0}E_j^{1\;1}E_{j+s_1}^{0\;0}+\Gamma_{0\;2}^{0\;2}E_j^{0\;0}E_{j+1}^{2\;2}+\Gamma_{2\;0}^{2\;0}E_j^{2\;2}E_{j+s_2}^{0\;0} \nonumber \\
&&\left.+\Gamma_{2\;1}^{2\;1}E_j^{2\;2}E_{j+s_2}^{1\;1}+\left(\Gamma_{0\;1}^{1\;0}+\Gamma_{2\;0}^{0\;2}\right)E_j^{1\;1}E_{j+s_1}^{2\;2} \right\} {\cal{P}},
\lb{e124}
\eea
where we chose $\Gamma_{0\;0}^{0\;0}=0$ but still we have $10$ free parameters! The particular choice $s_1=s_2=1$, $\Gamma_{2\;1}^{1\;2}=-\Gamma_{1\;2}^{1\;2}=\Gamma_{0\;1}^{1\;0}+\Gamma_{2\;0}^{0\;2}$, $\Gamma_{0\;1}^{1\;0}=-\Gamma_{1\;0}^{1\;0}$, $\Gamma_{2\;0}^{0\;2}=-\Gamma_{0\;2}^{0\;2}$, $\Gamma_{1\;1}^{1\;1}=\Gamma_{2\;2}^{2\;2}=\Gamma_{0\;1}^{0\;1}=\Gamma_{2\;0}^{2\;0}=0$, gives the Hamiltonian related to the stochastic problem of fully asymmetric diffusion of two kinds of particles, whose exact integrability  was obtained in \cite{popkov} through the dynamical matrix product  ansatz.  

        Let us return to the general case. The cyclic property of the trace in \rf{e122} and the commutation relations \rf{e89}, \rf{e115} and \rf{e119} give us
\bea
&&\mbox{Tr} \left[ Y_{k_1}^{(Q_1)}\cdots  Y_{k_n}^{(Q_n)}E^{L-n_1(s_1-1)-n_2(s_2-1)}\Omega_P \right] = e^{ik_j[L-n_1(s_1-1)-n_2(s_2-1)]}\nonumber \\
&&\times \sum_{Q_1',\ldots,Q_n'=1}^2 \langle Q_1,\ldots,Q_n|{\cal{T}}|Q_1',\ldots,Q_n'\rangle \mbox{Tr} \left[ Y_{k_1}^{(Q_1')}\cdots  Y_{k_n}^{(Q_n')}E^{L-n_1(s_1-1)-n_2(s_2-1)}\Omega_P \right],
\lb{e125}
\eea
where we have used the identity (see \rf{e99} and \rf{e111})
\beq
\sum_{Q_j'',Q_{j+1}''}S_{Q_j'\;Q_j''}^{Q_j\;Q_{j+1}''}(k_j,k_j)=-1,
\lb{e126}
\eeq
and
\bea
&&\langle Q_1,\ldots,Q_n|{\cal{T}}|Q_1',\ldots,Q_n'\rangle \nonumber \\
&&= \sum_{Q_1'',\ldots,Q_n''} \left\{ S_{Q_1'\;Q_1''}^{Q_1\;Q_2''}(k_1,k_j)\cdots S_{Q_j'\;Q_j''}^{Q_j\;Q_{j+1}''}(k_j,k_j)\cdots S_{Q_n'\;Q_n''}^{Q_n\;Q_1''}(k_n,k_j) e^{iP(s_1''-1)} \right\}
\lb{e127}
\eea
is the transfer matrix of a $2^n \times 2^n$-dimensional transfer matrix of an inhomogeneous $6$-vertex model (inhomogeneities $\{k_l \}$) with Boltzmann weights given by \rf{e99} and \rf{e111}. The $6$-vertex model is defined on a cylinder of transversal perimeter $n$ with a seam along its axis producing the twisted boundary conditions
\beq
S_{Q_n'\;Q_n''}^{Q_n\;Q_{n+1}''}(k_n,k)=S_{Q_n'\;Q_n''}^{Q_n\;Q_1''}(k_n,k)\phi(s_{Q_1''}),
\lb{e128}
\eeq
where
\beq
\phi(s)=e^{iP(s-1)},
\lb{e128p}
\eeq
and, as before, $P$ is the momentum of the eigenstate \rf{e86}. The relation \rf{e125} give us the constraints for the spectral parameters
\beq
e^{-ik_j(L+n-n_1s_1-n_2s_2)}=\Lambda(k_j,\{k_l \}) \;\;\; (j=1,\ldots,n), 
\lb{e129}
\eeq
where $\Lambda(k_j,\{k_l \})$ are the eigenvalues of the transfer matrix \rf{e127}. The condition \rf{e129} leads to the problem of evaluation the eigenvalues of the inhomogeneous transfer matrix \rf{e127}. This can be done through the algebraic Bethe  ansatz \cite{kulish} or the coordinate Bethe  ansatz (see \cite{alcbar4} and \cite{bjp4} for example), and we obtain
\beq
\Lambda(k_j,\{k_l \})=\phi(s_2)\prod_{l=1}^n S_{2\;2}^{2\;2}(k_l,k_j)\prod_{l=1}^{n_1}\frac{S_{2\;2}^{2\;2}(k_j,k_l^{(1)})}{S_{2\;1}^{2\;1}(k_j,k_l^{(1)})},
\lb{e130}
\eeq
where the unknown parameters $k_l^{(1)}$ ($l=1,\ldots,n_1$) are fixed by the $n_1$ coupled equations
\beq
\frac{\phi(s_1)}{\phi(s_2)}\prod_{l=1}^{n}\frac{S_{2\;1}^{2\;1}(k_l,k_j^{(1)})}{S_{2\;2}^{2\;2}(k_l,k_j^{(1)})}=\prod_{l=1}^{n_1}\frac{S_{2\;2}^{2\;2}(k_j^{(1)},k_l^{(1)})}{S_{1\;1}^{1\;1}(k_l^{(1)},k_j^{(1)})}\frac{S_{2\;1}^{2\;1}(k_l^{(1)},k_j^{(1)})}{S_{2\;1}^{2\;1}(k_j^{(1)},k_l^{(1)})},
\lb{e131}
\eeq
for $j=1,\ldots,n_1$. The equations \rf{e129}-\rf{e131} fix the spectral parameters $\{k_j;j=1,\ldots,n \}$ of the matrices $Y_{k_j}^{(Q)}$ introduced in our matrix product  ansatz.


\section{Models of spin-$\frac{3}{2}$ with two conservation laws.}

        Like in the previous section we now formulate our matrix product  ansatz for U($1$)$\otimes$U($1$) symmetric models describing the dynamics of two types of particles on the lattice, where the total number of particles of each type is conserved separately. However now, distinctly from the last section,  particles of distinct types may occupy the same site on the lattice (double occupancy). Models on this category are the spin-$\frac{3}{2}$ anisotropic Perck-Schultz model \cite{perkshultz}, the Essler-Korepin-Schoutens model \cite{essler2}, the Hubbard model \cite{lieb}, the Hamiltonian derived in \cite{ramos} from the $R$-matrix introduced in \cite{deguchi}, and the two-parameter integrable model introduced in \cite{alcbar2}.

	In order to define the Hamiltonians for these models on a lattice with $L$ sites let us attach to each lattice site a variable $Q_i$ ($i=1,\ldots,L$), that takes the values $Q_i=0$ if the site is empty, $Q_i=1,2$ if the site is   occupied by a particle of type $1$ or $2$, respectively, and $Q_i=3$ if the site have a double occupancy of particles of type $1$ and $2$. In the fermionic version of the models the particles of types $1$ and $2$ are the electrons with distinct spin polarizations. The most general model with nearest neighbor interactions and periodic boundary condition is given by
\bea
&&H^{U(1)\otimes U(1)}=-\sum_{j=1}^L H_{j,j+1} \nonumber \\
&&H_{j,j+1}=\sum_{\alpha \ne \beta =0}^3 \sum_{\gamma \ne \nu =0}^3 \Gamma_{\gamma\;\nu}^{\alpha\;\beta}E_j^{\gamma\;\alpha}E_{j+1}^{\nu\;\beta}+\sum_{\alpha =0}^3 \sum_{\beta=1}^3 \Gamma_{\alpha\;\beta}^{\alpha\;\beta}E_j^{\alpha\;\alpha}E_{j+1}^{\beta\;\beta},
\lb{f1}
\eea
where the coupling constants satisfy $\Gamma_{\gamma\;\nu}^{\alpha\;\beta}=0$ if $\alpha+\beta \ne \gamma+\nu$, $\Gamma_{\beta\;\gamma}^{\alpha\;\alpha}=\Gamma_{\alpha\;\alpha}^{\beta\;\gamma}=0$ unless $\alpha=\beta=\gamma$, and $E^{\alpha\;\beta}$ ($\alpha,\beta=0,1,2,3$) are the $4 \times 4$ Weyl matrices with $i,j$ elements $\left(E^{\alpha\;\beta} \right)_{i,j}=\delta_{\alpha,i}\delta_{\beta,j}$. Without loosing any generality we also chose hereafter $\Gamma_{0\;0}^{0\;0}=0$. The first and second summation in \rf{f1} account for the kinetic and static interactions. The U($1$)$\otimes$U($1$) symmetry and the translation symmetry, due to the periodic boundary condition of \rf{f1}, imply that the total number of particles $n_1,n_2$ ($0,1,\ldots$) of particles of types $1$ and $2$ and the momentum $P=\frac{2\pi l}{L}$ ($l=0,1,\ldots,L-1$) are conserved separately.

	We formulate a matrix product  ansatz for the eigenvectors $|\Psi_{n_1,n_2,P}\rangle$ of the eigenvalue equation
\beq
H^{U(1)\otimes U(1)}|\Psi_{n_1,n_2,P}\rangle=\varepsilon_{n_1,n_2}|\Psi_{n_1,n_2,P}\rangle
\lb{f2}
\eeq
in the eigensector ($n_1,n_2,P$). An arbitrary eigenfunction is given by
\beq
|\Psi_{n_1,n_2,P}\rangle=\sum_{\{Q \}}\sum_{\{x \}}f(x_1,Q_1;\ldots;x_n,Q_n)|x_1,Q_1;\ldots;x_n,Q_n\rangle
\lb{f3}
\eeq
where $|x_1,Q_1;\ldots;x_n,Q_n\rangle$ are the configurations whose particles of type $Q_i=1,2$ are located at $x_i=1,\ldots,L$ and $n=n_1+n_2$. The summation $\{Q \}=\{Q_1,\ldots,Q_n \}$ extends over all permutations of $n$ numbers $\{1,2 \}$ in which $n_1$ terms have the value $1$ and $n_2$ terms the value $2$. The summation $\{x \}=\{x_1,\ldots,x_n \}$ extends, for each permutation $\{Q \}$,  into the set of non-decreasing integers satisfying
\bea
&&(x_{i+1}-x_i) \ge 1 \;\;\; \mbox{if} \;\;\; Q_{i+1}=Q_i \nonumber \\
&&(x_{i+1}-x_i) \ge 0 \;\;\; \mbox{if} \;\;\; Q_{i+1}\ne Q_i, \;\;\; (i=1,\ldots,n-1).
\lb{f4}
\eea

	In order to formulate a matrix product  ansatz we associate to the 
	sites occupied by $Q_i=0,1$ and $2$ the matrices $E$, $Y^{(1)}$ and $Y^{(2)}$, respectively. In analogy to the results of section $3$ we associate to the double occupied sites ($Q_i=3$) the matrix product $Y^{(3)}=B^{(1)}E^{-1}B^{(2)}$. Our  ansatz asserts that the amplitudes corresponding to the configurations with no double occupied sites are given by the traces
\beq
f(x_1,Q_1;\ldots;x_n,Q_n)= \mbox{Tr} \left[E^{x_1-1}Y^{(Q_1)}E^{x_2-x_1-1}Y^{(Q_2)}\cdots E^{x_n-x_{n-1}-1}Y^{(Q_n)}E^{L-x_n}\Omega_P \right],
\lb{f5}
\eeq
while the amplitudes related with configurations with double occupancy at $x_{i+1}=x_i$ are given by the traces
\bea
&& f(x_1,Q_1;\ldots;x_i,1;x_i,2;\ldots;x_n,Q_n)= \mbox{Tr} \left[E^{x_1-1}Y^{(Q_1)}\right. \nonumber \\
&& \left. \cdots E^{x_i-x_{i-1}-1}B^{(1)}E^{-1}B^{(2)} 
 E^{x_{i+1}-x_i-1}\cdots E^{L-x_n}\Omega_P \right].
\lb{f6}
\eea
The general cases are given by generalizations of \rf{f5} and \rf{f6}.

	Similarly as in the previous sections the matrix $\Omega_P$ satisfying
\beq
E\Omega_P=e^{-iP}\Omega_PE, \;\;\; Y^{(Q)}\Omega_P=e^{-iP}\Omega_PY^{(Q)}, \;\;\; B^{(Q)}\Omega_P=e^{-iP}\Omega_PB^{(Q)} \;\;\; (Q=1,2)
\lb{f6p}
\eeq
ensures (see \rf{e6}) the momentum $P$ of the eigenvector $|\Psi_{n_1,n_2,P}\rangle$.

	As in the previous sections in order to satisfy the eigenvalue equation \rf{f2} we identify $Y^{(Q)}$ and $B^{(Q)}$ ($Q=1,2$) as composed by $n$ spectral parameter dependent matrices \footnote{Similarly as the solutions \rf{e102} in section $4$ there exist special solutions, that for brevity we do not consider here, where the matrices  $Y^{(1)}$, $B^{(1)}$ and $Y^{(2)}$, $B^{(2)}$ are composed by the distinct sets of spectral parameters $\{k_1,\ldots,k_{n_1} \}$ and $\{k_{n_1},\ldots,k_{n_1+n_2} \}$, respectively.}
\beq
Y^{(Q)}=\sum_{j=1}^nY_{k_j}^{(Q)}E, \;\;\; B^{(Q)}=\sum_{j=1}^nB_{k_j}^{(Q)}E (Q=1,2),
\lb{f7}
\eeq
satisfying the commutation relations
\bea
&&Y_{k_j}^{(Q)}E=e^{ik_j}EY_{k_j}^{(Q)}, \;\;\; B_{k_j}^{(Q)}E=e^{ik_j}EB_{k_j}^{(Q)}, \nonumber \\
&&\left[Y_{k_j}^{(Q)}, \Omega_P \right]=\left[B_{k_j}^{(Q)}, \Omega_P \right]= Y_{k_j}^{(Q)}Y_{k_j}^{(Q')}=B_{k_j}^{(Q)}B_{k_j}^{(Q')}=0 \;\;\; (Q,Q'=1,2),
\lb{f8}
\eea
where $k_j$ ($j=1,\ldots,n$) are unknown complex spectral parameters.

	Let us consider initially the sectors with small values of $n$.

{\it {\bf  n = 1.}}
We have distinct equations depending on the type $Q=1,2$ of the particle. The eigenvalue equation \rf{f2} give us, for $Q=1,2$,
\bea
\varepsilon^{(Q)}\mbox{Tr} \left[E^{x-1}Y^{(Q)}E^{L-x}\Omega_P \right]=&-&\Gamma_{0\;Q}^{Q\;0}\mbox{Tr} \left[E^{x-2}Y^{(Q)}E^{L-x+1}\Omega_P \right]-\Gamma_{Q\;0}^{0\;Q}\mbox{Tr} \left[E^xY^{(Q)}E^{L-x-1}\Omega_P \right] \nonumber \\
&-&\left(\Gamma_{0\;Q}^{0\;Q}+\Gamma_{Q\;0}^{Q\;0} \right)\mbox{Tr} \left[E^{x-1}Y^{(Q)}E^{L-x}\Omega_P \right].
\lb{f9}
\eea
Inserting \rf{f7} with $n=1$ and using \rf{f8} we obtain the energy and momentum as a function of the spectral parameters 
\bea
&&\varepsilon^{(Q)}(k)=-\left(\Gamma_{0\;Q}^{Q\;0}e^{-ik}+\Gamma_{Q\;0}^{0\;Q}e^{ik}+\Gamma_{0\;Q}^{0\;Q}+\Gamma_{Q\;0}^{Q\;0} \right) \;\;\; (Q=1,2), \nonumber \\
&&P=k=\frac{2\pi}{L}l \;\;\; (l=0,1,\ldots,L-1).
\lb{f10}
\eea

{\it {\bf  n = 2.}}
For two particles of type $Q_1$, and $Q_2$ ($Q_1,Q_2=1,2$) located at $x_1$ and $x_2$ the eigenvalue equation \rf{f2} produce distinct relations. The components where the particles are not at ``colliding'' positions, $x_{i+1}>x_i+1$, give us the relation
\bea
&&\varepsilon^{(Q_1,Q_2)}\mbox{Tr} \left[E^{x_1-1}Y^{(Q_1)}E^{x_2-x_1-1}Y^{(Q_2)}E^{L-x_2}\Omega_P \right]= \nonumber \\
&&-\Gamma_{0\;Q_1}^{Q_1\;0}\mbox{Tr} \left[E^{x_1-2}Y^{(Q_1)}E^{x_2-x_1}Y^{(Q_2)}E^{L-x_2}\Omega_P \right]-\Gamma_{Q_1\;0}^{0\;Q_1}\mbox{Tr} \left[E^{x_1}Y^{(Q_1)}E^{x_2-x_1-2}Y^{(Q_2)}E^{L-x_2}\Omega_P \right] \nonumber \\
&&-\Gamma_{0\;Q_2}^{Q_2\;0}\mbox{Tr} \left[E^{x_1-1}Y^{(Q_1)}E^{x_2-x_1-2}Y^{(Q_2)}E^{L-x_2+1}\Omega_P \right] \nonumber \\
&&-\Gamma_{Q_2\;0}^{0\;Q_2}\mbox{Tr} \left[E^{x_1-1}Y^{(Q_1)}E^{x_2-x_1}Y^{(Q_2)}E^{L-x_2-1}\Omega_P \right] \nonumber \\
&&-\left(\Gamma_{0\;Q_1}^{0\;Q_1}+\Gamma_{Q_1\;0}^{Q_1\;0}+\Gamma_{0\;Q_2}^{0\;Q_2}+\Gamma_{Q_2\;0}^{Q_2\;0} \right)\mbox{Tr} \left[E^{x_1-1}Y^{(Q_1)}E^{x_2-x_1-1}Y^{(Q_2)}E^{L-x_2}\Omega_P \right].
\lb{f11}
\eea
Inserting \rf{f5} with $n=2$ into this last expression and using \rf{f8} we obtain a solution provide the coupling constants in \rf{f1} satisfy the constraint
\beq
\Gamma_{1\;0}^{0\;1}=\Gamma_{2\;0}^{0\;2}, \;\;\; \Gamma_{0\;1}^{1\;0}=\Gamma_{0\;2}^{2\;0}.
\lb{f12}
\eeq
The energy and momentum in terms of the spectral parameters $k_1$ and $k_2$ are given by
\beq
\varepsilon^{(Q_1,Q_2)}=\varepsilon^{(Q_1)}(k_1)+\varepsilon^{(Q_2)}(k_2), \;\;\; P=k_1+k_2,
\lb{f13}
\eeq
where $\varepsilon^{(Q)}(k)$ is given by \rf{f10}.

	In order to proceed it is better to consider separately the cases where $Q_1=Q_2$ and $Q_1 \ne Q_2$. If $Q_1=Q_2$, \rf{f2} applied to the amplitudes related to the configurations where $x_1=x$, $x_2=x+1$ give us the relation 
\bea
&&\varepsilon^{(Q,Q)}\mbox{Tr}\left[E^{x-1}Y^{(Q)}Y^{(Q)}E^{L-x-1}\Omega_P\right]=-\Gamma_{0\;Q}^{Q\;0}\mbox{Tr}\left[E^{x-2}Y^{(Q)}EY^{(Q)}E^{L-x-1}\Omega_P\right] \nonumber \\
&&-\Gamma_{Q\;0}^{0\;Q}\mbox{Tr}\left[E^{x-1}Y^{(Q)}EY^{(Q)}E^{L-x-2}\Omega_P\right]-\left(\Gamma_{0\;Q}^{0\;Q}+\Gamma_{Q\;0}^{Q\;0}+\Gamma_{Q\;Q}^{Q\;Q}\right)\mbox{Tr}\left[E^{x-1}Y^{(Q)}Y^{(Q)}E^{L-x-1}\Omega_P\right]. \nonumber \\
\lb{f14}
\eea
Inserting \rf{f7} and \rf{f13} in \rf{f14} and using \rf{f8} we obtain the algebraic relations among the spectral parameter matrices 
\beq
Y_{k_j}^{(Q)}Y_{k_l}^{(Q)}=S_{Q\;Q}^{Q\;Q}(k_j,k_l)Y_{k_l}^{(Q)}Y_{k_j}^{(Q)}, \;\;\; (Q=1,2), 
\lb{f15}
\eeq
where 
\beq
S_{Q\;Q}^{Q\;Q}(k_j,k_l)=-\frac{\Gamma_{Q\;0}^{0\;Q}e^{i(k_j+k_l)}+\Gamma_{0\;Q}^{Q\;0}+\left(\Gamma_{0\;Q}^{0\;Q}+\Gamma_{Q\;0}^{Q\;0}-\Gamma_{Q\;Q}^{Q\;Q}\right)e^{ik_j}}{\Gamma_{Q\;0}^{0\;Q}e^{i(k_j+k_l)}+\Gamma_{0\;Q}^{Q\;0}+\left(\Gamma_{0\;Q}^{0\;Q}+\Gamma_{Q\;0}^{Q\;0}-\Gamma_{Q\;Q}^{Q\;Q}\right)e^{ik_l}} \;\;\; (Q=1,2).
\lb{f15p}
\eeq

	If $Q_1 \ne Q_2$ distinct relations merge from the configurations where $x_1=x_2=x$ and $x_1=x$, $x_2=x+1$. For $x_1=x_2=x$ we have
\bea
&\varepsilon^{(1,2)}\mbox{Tr}\left[E^{x-1}B^{(1)}E^{-1}B^{(2)}E^{L-x}\Omega_P\right]=-\Gamma_{0\;3}^{1\;2}\mbox{Tr}\left[E^{x-2}Y^{(1)}Y^{(2)}E^{L-x}\Omega_P\right] \nonumber \\
&-\Gamma_{3\;0}^{2\;1}\mbox{Tr}\left[E^{x-1}Y^{(2)}Y^{(1)}E^{L-x-1}\Omega_P\right]-\Gamma_{0\;3}^{2\;1}\mbox{Tr}\left[E^{x-2}Y^{(2)}Y^{(1)}E^{L-x}\Omega_P\right] \nonumber \\
&-\Gamma_{3\;0}^{1\;2}\mbox{Tr}\left[E^{x-1}Y^{(1)}Y^{(2)}E^{L-x-1}\Omega_P\right]-\Gamma_{0\;3}^{3\;0}\mbox{Tr}\left[E^{x-2}B^{(1)}E^{-1}B^{(2)}E^{L-x+1}\Omega_P\right] \nonumber \\
&-\Gamma_{3\;0}^{0\;3}\mbox{Tr}\left[E^xB^{(1)}E^{-1}B^{(2)}E^{L-x-1}\Omega_P\right]-\left(\Gamma_{0\;3}^{0\;3}+\Gamma_{3\;0}^{3\;0}\right)\mbox{Tr}\left[E^{x-1}B^{(1)}E^{-1}B^{(2)}E^{L-x}\Omega_P\right],
\lb{f16}
\eea
while for $x_1=x$, $x_2=x+1$ we have for $Q_1 \ne Q_2$ ($Q_1,Q_2=1,2$)
\bea
&\varepsilon^{(Q_1,Q_2)}\mbox{Tr}\left[E^{x-1}Y^{(Q_1)}Y^{(Q_2)}E^{L-x-1}\Omega_P\right]=-\Gamma_{0\;Q_1}^{Q_1\;0}\mbox{Tr}\left[E^{x-2}Y^{(Q_1)}EY^{(Q_2)}E^{L-x-1}\Omega_P\right] \nonumber \\
&-\Gamma_{Q_2\;0}^{0\;Q_2}\mbox{Tr}\left[E^{x-1}Y^{(Q_1)}EY^{(Q_2)}E^{L-x-2}\Omega_P\right]-\Gamma_{Q_1\;Q_2}^{3\;0}\mbox{Tr}\left[E^{x-1}B^{(1)}E^{-1}B^{(2)}E^{L-x}\Omega_P\right] \nonumber \\
&-\Gamma_{Q_1\;Q_2}^{0\;3}\mbox{Tr}\left[E^xB^{(1)}E^{-1}B^{(2)}E^{L-x-1}\Omega_P\right]-\Gamma_{Q_1\;Q_2}^{Q_2\;Q_1}\mbox{Tr}\left[E^{x-1}Y^{(Q_2)}Y^{(Q_1)}E^{L-x-1}\Omega_P\right] \nonumber \\
&-\left(\Gamma_{0\;Q_1}^{0\;Q_1}+\Gamma_{Q_2\;0}^{Q_2\;0}-\Gamma_{Q_1\;Q_2}^{Q_1\;Q_2}\right)\mbox{Tr}\left[E^{x-1}Y^{(Q_1)}Y^{(Q_2)}E^{L-x-1}\Omega_P\right].
\lb{f17}
\eea
Inserting \rf{f7} and \rf{f13} in \rf{f16} and \rf{f17} and using \rf{f8} we obtain the algebraic relations
\beq
\sum_{j,l=1\;j\ne l}^2C_0(k_j,k_l)B_{k_j}^{(1)}B_{k_l}^{(2)}=-\sum_{j,l=1\;j\ne l}^2e^{ik_l}\left[C_1^{'}(k_j,k_l)Y_{k_j}^{(1)}Y_{k_l}^{(2)}+C_2^{'}(k_j,k_l)Y_{k_j}^{(2)}Y_{k_l}^{(1)}\right],
\lb{f18}
\eeq
\bea
&&\sum_{j,l=1\;j\ne l}^2 C_Q(k_j,k_l)B_{k_j}^{(1)}B_{k_l}^{(2)}=\sum_{j,l=1\;j\ne l}^2 \left\{ \left[ D(k_j,k_l)-\left(\Gamma_{Q_1\;0}^{Q_1\;0}+\Gamma_{0\;Q_2}^{0\;Q_2}-\Gamma_{Q_1\;Q_2}^{Q_1\;Q_2} \right)e^{ik_l} \right]Y_{k_j}^{(Q_1)}Y_{k_l}^{(Q_2)} \right. \nonumber \\
&&\left.+\Gamma_{Q_1\;Q_2}^{Q_2\;Q_1}e^{ik_l}Y_{k_j}^{(Q_2)}Y_{k_l}^{(Q_1)}\right\}, \;\;\; Q_1 \ne Q_2 \;\;\; (Q_1=1,2),
\lb{f19}
\eea
where $C_0$, $C_1^{'}$, $C_2^{'}$, $C_1$, $C_2$ and $D$ are the symmetric functions
\bea
&&C_0(k_j,k_l)=-\Gamma_{0\;1}^{1\;0}\left(e^{ik_j}+e^{ik_l} \right)-\Gamma_{1\;0}^{0\;1}e^{i(k_j+k_l)}\left(e^{ik_j}+e^{ik_l} \right) \nonumber \\
&&-\left(\Gamma_{1\;0}^{1\;0}+\Gamma_{0\;1}^{0\;1}+\Gamma_{2\;0}^{2\;0}+\Gamma_{0\;2}^{0\;2}-\Gamma_{0\;3}^{0\;3}-\Gamma_{3\;0}^{3\;0} \right)e^{i(k_j+k_l)}+ \Gamma_{0\;3}^{3\;0}+\Gamma_{3\;0}^{0\;3}e^{i2(k_j+k_l)}, \nonumber \\
&&C_1(k_j,k_l)=-\Gamma_{1\;2}^{3\;0}-\Gamma_{1\;2}^{0\;3}e^{i(k_j+k_l)}, \;\; C_1^{'}(k_j,k_l)=\Gamma_{0\;3}^{1\;2}+\Gamma_{3\;0}^{1\;2}e^{i(k_j+k_l)}, \nonumber \\
&&C_2(k_j,k_l)=-\Gamma_{2\;1}^{3\;0}-\Gamma_{2\;1}^{0\;3}e^{i(k_j+k_l)}, \;\; C_2^{'}(k_j,k_l)=\Gamma_{0\;3}^{2\;1}+\Gamma_{3\;0}^{2\;1}e^{i(k_j+k_l)}, \nonumber \\
&&D(k_j,k_l)=-\Gamma_{0\;1}^{1\;0}-\Gamma_{1\;0}^{0\;1}e^{i(k_j+k_l)}.
\lb{f20}
\eea
Multiplying the two relations in \rf{f19} by $C_0(k_j,k_l)$ and using \rf{f18} we obtain the algebraic relations expressed in the matrix notation
\beq
\label{f21}
\sum_{j\neq l=1}^2
$$\left[ \matrix{  \alpha_{j,l}  &
\beta_{j,l} \cr
  \gamma_{j,l} & \delta_{j,l}}
        \right]
\left[ \matrix{ Y_{k_j}^{(1)}Y_{k_l}^{(2)} \cr
  Y_{k_j}^{(2)}Y_{k_l}^{(1)}     }
\right] = 0.$$
\eeq

where
\bea
&&\alpha_{j,l}=-C_1(k_j,k_l)C_1^{'}(k_j,k_l)e^{ik_l}-C_0(k_j,k_l)\left[D(k_j,k_l)-\left(\Gamma_{1\;0}^{1\;0}+\Gamma_{0\;2}^{0\;2}-\Gamma_{1\;2}^{1\;2} \right)e^{ik_l} \right] \nonumber \\
&&\beta_{j,l}=-\left[C_1(k_j,k_l)C_2^{'}(k_j,k_l)-C_0(k_j,k_l)\Gamma_{1\;2}^{2\;1} \right]e^{ik_l} \nonumber \\
&&\gamma_{j,l}=-\left[C_2(k_j,k_l)C_1^{'}(k_j,k_l)-C_0(k_j,k_l)\Gamma_{2\;1}^{1\;2} \right]e^{ik_l} \nonumber \\
&&\delta_{j,l}=-C_2(k_j,k_l)C_2^{'}(k_j,k_l)e^{ik_l}-C_0(k_j,k_l)\left[D(k_j,k_l)-\left(\Gamma_{0\;1}^{0\;1}+\Gamma_{2\;0}^{2\;0}-\Gamma_{2\;1}^{2\;1} \right)e^{ik_l} \right].
\lb{f22}
\eea
This last equation can be rearranged straightforwardly, giving us
\beq
Y_{k_l}^{(Q_1)}Y_{k_m}^{(Q_2)}=S_{Q_1\;Q_2}^{Q_1\;Q_2}(k_l,k_m)Y_{k_m}^{(Q_2)}Y_{k_l}^{(Q_1)}+S_{Q_2\;Q_1}^{Q_1\;Q_2}(k_l,k_m)Y_{k_m}^{(Q_1)}Y_{k_l}^{(Q_2)}
\lb{f23}
\eeq
where
\bea
&&S_{2\;1}^{1\;2}(k_j,k_l)=-\left(1+\frac{\delta_{j,l}\left(\alpha_{l,j}-\alpha_{j,l} \right)-\beta_{j,l}\left(\gamma_{l,j}-\gamma_{j,l} \right)}{\alpha_{j,l}\delta_{j,l}-\beta_{j,l}\gamma_{j,l}} \right), \nonumber \\
&&S_{1\;2}^{1\;2}(k_j,k_l)=-\frac{\delta_{j,l}\beta_{l,j}-\beta_{j,l}\delta_{l,j}}{\alpha_{j,l}\delta_{j,l}-\beta_{j,l}\gamma_{j,l}}, \nonumber \\
&&S_{1\;2}^{2\;1}(k_j,k_l)=-\left(1+\frac{\alpha_{j,l}\left(\delta_{l,j}-\delta_{j,l} \right)-\gamma_{j,l}\left(\beta_{l,j}-\beta_{j,l} \right)}{\alpha_{j,l}\delta_{j,l}-\beta_{j,l}\gamma_{j,l}} \right), \nonumber \\
&&S_{2\;1}^{2\;1}(k_j,k_l)=-\frac{\alpha_{j,l}\gamma_{l,j}-\gamma_{j,l}\alpha_{l,j}}{\alpha_{j,l}\delta_{j,l}-\beta_{j,l}\gamma_{j,l}}.
\lb{f24}
\eea
Multiplying \rf{f19} with $Q_1=1$ by $C_1^{'}$ and \rf{f18} by $\Gamma_{2\;1}^{1\;2}$ and subtracting the obtained expressions we obtain a relation that express the spectral parameter matrices $B_k^{(Q)}$ in terms of $Y_k^{(Q)}$, i. e.,
\beq
B_{k_j}^{(1)}B_{k_l}^{(2)}=\frac{D_3(k_j,k_l)}{\rho(k_j,k_l)}Y_{k_j}^{(1)}Y_{k_l}^{(2)} \;\;\; (j \ne l),
\lb{f24p}
\eeq
where
\bea
&&D_3(k_j,k_l)=D(k_j,k_l)-\left(\Gamma_{1\;0}^{1\;0}-\Gamma_{1\;2}^{1\;2} \right)C_2^{'}(k_j,k_l)-C_1^{'}(k_j,k_l)\Gamma_{1\;2}^{2\;1}e^{ik_l}, \nonumber \\
&&\rho(k_j,k_l)=-\beta_{j,l}e^{-ik_l}=C_1(k_j,k_l)C_2^{'}(k_j,k_l)-C_0(k_j,k_l)\Gamma_{1\;2}^{2\;1}.
\lb{f24pp}
\eea
Since the matrices $B_{k_j}^{(Q)}$ only appear in the pairs $B_{k_j}^{(1)}B_{k_j}^{(2)}$ in our matrix product  ansatz \rf{f5}-\rf{f6} the relation \rf{f24p} enable us to express the amplitudes only in terms of the spectral parameter matrices $Y_{k_j}^{(Q)}$ ($Q=1,2$). Consequently \rf{f23}, \rf{f24} together with \rf{f8} give us the complete algebraic relations for the matrices appearing in our matrix product  ansatz in the sector with $n=2$ particles.

	The spectral parameters $k_1$ and $k_2$, as in the previous sections, are fixed by the cyclic property of the trace defining the amplitudes. Instead of producing the equations fixing $\{k_j \}$ for $n=2$ let us consider the cases of general values of $n$.

{\it {\bf  General n.}}
We now consider the case of an arbitrary number $n_1$, $n_2$ of particles of type $1$ and $2$ ($n=n_1+n_2$). The eigenvalue equation \rf{f2} when applied to the components of $|\Psi_{n_1,n_2,P}\rangle$ with no particles at ``colliding'' positions ($x_{i+1}>x_i+1$, $i=1,\ldots,n$) give us a generalization of \rf{f11} that is solved by giving us the energy and momentum as a function of the spectral parameters 
\bea
&&\varepsilon_{n_1,n_2}=\sum_{j=1}^{n_1}\varepsilon^{(1)}(k_j)+
\sum_{j=n_1+1}^n\varepsilon^{(2)}(k_j)=
\sum_{j=1}^n\left(\Gamma_{0\;1}^{1\;0}e^{-ik_j}+\Gamma_{1\;0}^{0\;1}e^{ik_j} \right)
\nonumber \\
&& +n_1\left(\Gamma_{1\;0}^{1\;0}+\Gamma_{0\;1}^{0\;1} \right)+n_2\left(\Gamma_{2\;0}^{2\;0}+
\Gamma_{0\;2}^{0\;2} \right), 
 \; \; \;  
P=\sum_{j=1}^nk_j.
\lb{f25}
\eea

	The amplitudes of $|\Psi_{n_1,n_2,P}\rangle$ where a pair of particles of type $Q_1$ and $Q_2$ are located at the colliding positions $x_{i+1}=x_i$ or $x_{i+1}=x_i+1$ give us algebraic relations that are generalizations of \rf{f15} and \rf{f23},
\bea
&&Y_{k_j}^{(Q_1)}Y_{k_l}^{(Q_2)}=
\sum_{Q_1',Q_2'=1}^2S_{Q_2'\;Q_1'}^{Q_1\;Q_2}(k_j,k_l)Y_{k_l}^{(Q_1')}Y_{k_j}^{(Q_2')} \;\;\; 
(k_j \ne k_l), \nonumber \\
&&Y_{k_j}^{(Q_1)}Y_{k_j}^{(Q_2)}=0 \;\;\; (j,l=1,\ldots,n),
\lb{f26}
\eea
where the nonzero elements of the $S$-matrix in \rf{f26} are those given in \rf{f15p} and \rf{f24}. Moreover any pair of matrices $B_{k_j}^{(1)}B_{k_l}^{(2)}$ ($j \ne l$; $j,l=1,\ldots ,n$) are expressed in terms of $Y_{k_j}^{(1)}Y_{k_l}^{(2)}$ by \rf{f24p}

	For arbitrary amplitudes the matrix product  ansatz \rf{f5} and \rf{f6}, after using \rf{f24p}, contains a product of $n$ spectral matrices $\{Y_{k_j}^{(Q)} \}$. The  ansatz will be valid only if the algebraic relations among these matrices give us a unique relation among these products. This imply that products like $\cdots Y_{k_1}^{(Q_1)}Y_{k_2}^{(Q_2)}Y_{k_3}^{(Q_3)} \cdots $ and $\cdots Y_{k_3}^{(Q_3)}Y_{k_2}^{(Q_2)}Y_{k_1}^{(Q_1)} \cdots $ should be uniquely related. Similarly as discussed in section $4$ these products can be related by applying the commutation relations in distinct ways. Performing the commutation relations in the order $Q_1Q_2Q_3 \rightarrow Q_2Q_1Q_3 \rightarrow Q_2Q_3Q_1 \rightarrow Q_3Q_2Q_1$ or $Q_1Q_2Q_3 \rightarrow Q_1Q_3Q_2 \rightarrow Q_3Q_1Q_2 \rightarrow Q_3Q_2Q_1$ will impose constraints among the structure constants $S_{Q_1\;Q_2}^{Q_1'\;Q_2'}$ of the algebraic relations \rf{f15p} and \rf{f24}, namely,
\bea
&&\sum_{Q,Q',Q''=1}^2 S_{Q\;Q'}^{Q_1\;Q_1'}(k_1,k_2)S_{Q_2\;Q''}^{Q\;Q_1''}(k_1,k_3)S_{Q_2'\;Q_2''}^{Q'\;Q''}(k_2,k_3) \nonumber \\
&&=\sum_{Q,Q',Q''=1}^2 S_{Q'\;Q''}^{Q_1'\;Q_1''}(k_2,k_3)S_{Q_1\;Q_2''}^{Q_1\;Q''}(k_1,k_3)S_{Q_2\;Q_2'}^{Q\;Q'}(k_1,k_2),
\lb{f27}
\eea
for $Q_1,Q_1',Q_1'',Q_2,Q_2',Q_2''=1,2$. These constraints are just the Yang-Baxter relations \cite{yang,baxter} of the $S$-matrix defined in \rf{f15p} and \rf{f24}. The Yang-Baxter relations \rf{f27} imply the associativity of the algebra of the matrices $\{Y_{k_j}^{(Q)} \}$. These relations will impose severe constraints among the coupling constants $\Gamma_{n\;o}^{l\;m}$ of our general Hamiltonian \rf{f1}. However the Yang-Baxter relations \rf{f27} are not enough to ensure that the eigenfunctions $|\Psi_{n_1,n_2,P}\rangle$ are given by our matrix product  ansatz, since new relations among the spectral parameters matrices $\{Y_{k_j}^{(Q)} \}$ happens when we have on the lattice $3$ or $4$ particles at matching conditions. These new relations happen because within the range of the nearest neighbor interactions of the Hamiltonian \rf{f1} we may have up to $4$ particles. The eigenvalue equation \rf{f2} applied to the amplitudes where we have a particle $Q$ at $x_1=x$ and a pair at $x_2=x+1$, and to the amplitude with a particle $Q$ at $x_1=x+1$ and a pair at $x$ give us, respectively, the equations
\bea
&&\left(\varepsilon^{(Q)}+\varepsilon^{(1)}+\varepsilon^{(2)}+\Gamma_{0\;Q}^{0\;Q}+\Gamma_{Q\;3}^{Q\;3}+\Gamma_{3\;0}^{3\;0} \right) \mbox{Tr} \left[E^{x-1}Y^{(Q)}B^{(1)}E^{-1}B^{(2)}E^{L-x-1}\Omega_P \right]= \nonumber \\
&&-\Gamma_{0\;Q}^{Q\;0}\mbox{Tr} \left[E^{x-2}Y^{(Q)}EB^{(1)}E^{-1}B^{(2)}E^{L-x-1}\Omega_P \right]-\Gamma_{3\;\;\;0}^{Q'\;Q}\mbox{Tr} \left[E^{x-1}Y^{(Q)}Y^{(Q')}Y^{(Q)}E^{L-x-2}\Omega_P \right] \nonumber \\
&&-\Gamma_{3\;\;0}^{Q\;Q'}\mbox{Tr} \left[E^{x-1}Y^{(Q)}Y^{(Q)}Y^{(Q')}E^{L-x-2}\Omega_P \right]-\Gamma_{3\;0}^{0\;3}\mbox{Tr} \left[E^{x-1}Y^{(Q)}EB^{(1)}E^{-1}B^{(2)}E^{L-x-2}\Omega_P \right] \nonumber \\
&&-\Gamma_{Q\;3}^{3\;Q}\mbox{Tr} \left[E^{x-1}B^{(1)}E^{-1}B^{(2)}Y^{(Q)}E^{L-x-1}\Omega_P \right], \;\;\; Q \ne Q' \;\;\; (Q,Q'=1,2),
\lb{f28}
\eea
\bea
&&\left(\varepsilon^{(Q)}+\varepsilon^{(1)}+\varepsilon^{(2)}+\Gamma_{0\;3}^{0\;3}+\Gamma_{3\;Q}^{3\;Q}+\Gamma_{Q\;0}^{Q\;0} \right) \mbox{Tr} \left[E^{x-1}B^{(1)}E^{-1}B^{(2)}Y^{(Q)}E^{L-x-1}\Omega_P \right]= \nonumber \\
&&-\Gamma_{Q\;0}^{0\;Q}\mbox{Tr} \left[E^{x-1}B^{(1)}E^{-1}B^{(2)}EY^{(Q)}E^{L-x-2}\Omega_P \right]-\Gamma_{0\;\;\;3}^{Q'\;Q}\mbox{Tr} \left[E^{x-2}Y^{(Q')}Y^{(Q)}Y^{(Q)}E^{L-x-1}\Omega_P \right] \nonumber \\
&&-\Gamma_{0\;\;3}^{Q\;Q'}\mbox{Tr} \left[E^{x-2}Y^{(Q)}Y^{(Q')}Y^{(Q)}E^{L-x-1}\Omega_P \right]-\Gamma_{0\;3}^{3\;0}\mbox{Tr} \left[E^{x-2}B^{(1)}E^{-1}B^{(2)}EY^{(Q)}E^{L-x-1}\Omega_P \right] \nonumber \\
&&-\Gamma_{3\;Q}^{Q\;3}\mbox{Tr} \left[E^{x-1}Y^{(Q)}B^{(1)}E^{-1}B^{(2)}E^{L-x-1}\Omega_P \right], \;\;\; Q \ne Q' \;\;\; (Q,Q'=1,2).
\lb{f29}
\eea
Inserting \rf{f7} and \rf{f25} into these last equations, and using the generalizations of \rf{f18} to simplify the obtained expressions we obtain the four algebraic relations
\bea
&&\sum_{j,l,m=1}^3 \left\{ D^{(3,Q)}(k_j,k_l,k_m)Y_{k_j}^{(Q)}B_{k_l}^{(1)}B_{k_m}^{(2)}\right. \nonumber \\
&&\left. -e^{ik_m}\left(\Gamma_{Q\;3}^{3\;Q}B_{k_j}^{(1)}B_{k_l}^{(2)}Y_{k_m}^{(Q)}-\Gamma_{0\;\;\;3}^{Q'\;Q}Y_{k_j}^{(Q)}Y_{k_l}^{(Q')}Y_{k_m}^{(Q)}-\Gamma_{0\;\;3}^{Q\;Q'}Y_{k_j}^{(Q)}Y_{k_l}^{(Q)}Y_{k_m}^{(Q')} \right) \right\}=0, \nonumber \\
&&\sum_{j,l,m=1}^3 \left\{ D^{(Q,3)}(k_j,k_l,k_m)B_{k_j}^{(1)}B_{k_l}^{(2)}Y_{k_m}^{(Q)}\right. \nonumber \\
&&\left. -e^{i(k_l+k_m)}\left(\Gamma_{3\;Q}^{Q\;3}Y_{k_j}^{(Q)}B_{k_l}^{(1)}B_{k_m}^{(2)}-\Gamma_{3\;\;\;0}^{Q'\;Q}Y_{k_j}^{(Q')}Y_{k_l}^{(Q)}Y_{k_m}^{(Q)}-\Gamma_{3\;\;0}^{Q\;Q'}Y_{k_j}^{(Q)}Y_{k_l}^{(Q')}Y_{k_m}^{(Q)} \right) \right\}=0,
\lb{f30}
\eea
where $Q,Q'=1,2$ or $Q,Q'=21$, and
\bea
&&D^{(3,Q)}(k_j,k_l,k_m)=\Gamma_{Q\;0}^{0\;Q}e^{i(k_j+k_l+k_m)}+\left(\Gamma_{Q\;0}^{Q\;0}+\Gamma_{0\;3}^{0\;3}-\Gamma_{Q\;3}^{Q\;3} \right)e^{i(k_l+k_m)}+\Gamma_{0\;3}^{3\;0}, \nonumber \\
&&D^{(Q,3)}(k_j,k_l,k_m)=\Gamma_{3\;0}^{0\;3}e^{i(k_j+k_l+k_m)}+\left(\Gamma_{0\;Q}^{0\;Q}+\Gamma_{3\;0}^{3\;0}-\Gamma_{3\;Q}^{3\;Q} \right)e^{ik_m}+\Gamma_{0\;Q}^{Q\;0}.
\lb{f31}
\eea
In the case of $n=4$ particles the new relation comes from the amplitudes corresponding to the configurations where we have two pairs of particles at sites $x$ and $x+1$. The eigenvalue equation \rf{f2}, when applied to these amplitudes give us 
\bea
&&\left(2\varepsilon^{(1)}+2\varepsilon^{(2)}+\Gamma_{0\;3}^{0\;3}+\Gamma_{3\;3}^{3\;3}+\Gamma_{3\;0}^{3\;0} \right) \mbox{Tr} \left[E^{x-1}B^{(1)}E^{-1}B^{(2)}B^{(1)}E^{-1}B^{(2)}E^{L-x-1}\Omega_P \right]= \nonumber \\
&&-\Gamma_{0\;3}^{2\;1}\mbox{Tr} \left[E^{x-2}Y^{(2)}Y^{(1)}B^{(1)}E^{-1}B^{(2)}E^{L-x-1}\Omega_P \right] \nonumber \\
&&-\Gamma_{0\;3}^{1\;2}\mbox{Tr} \left[E^{x-2}Y^{(1)}Y^{(2)}B^{(1)}E^{-1}B^{(2)}E^{L-x-1}\Omega_P \right] \nonumber \\
&&-\Gamma_{3\;0}^{1\;2}\mbox{Tr} \left[E^{x-1}B^{(1)}E^{-1}B^{(2)}Y^{(1)}Y^{(2)}E^{L-x-2}\Omega_P \right] \nonumber \\
&&-\Gamma_{3\;0}^{2\;1}\mbox{Tr} \left[E^{x-1}B^{(1)}E^{-1}B^{(2)}Y^{(2)}Y^{(1)}E^{L-x-2}\Omega_P \right] \nonumber \\
&&-\Gamma_{0\;3}^{3\;0}\mbox{Tr} \left[E^{x-2}B^{(1)}E^{-1}B^{(2)}EB^{(1)}E^{-1}B^{(2)}E^{L-x-1}\Omega_P \right] \nonumber \\
&&-\Gamma_{3\;0}^{0\;3}\mbox{Tr} \left[E^{x-1}B^{(1)}E^{-1}B^{(2)}EB^{(1)}E^{-1}B^{(2)}E^{L-x-2}\Omega_P \right].
\lb{f32}
\eea
Inserting \rf{f7} and \rf{f25} into this last relation and using the generalization of \rf{f18} to simplify the resulting expression we obtain 
\bea
&&\sum_{j,l,m,o=1}^4 \left\{ D^{(3,3)}(k_j,k_l,k_m,k_o)B_{k_j}^{(1)}B_{k_l}^{(2)}B_{k_m}^{(1)}B_{k_o}^{(2)}+e^{i(k_l+k_m+k_o)}\left(\Gamma_{3\;0}^{2\;1}Y_{k_j}^{(2)}Y_{k_l}^{(1)} \right. \right. \nonumber \\
&&\left. \left. +\Gamma_{3\;0}^{1\;2}Y_{k_j}^{(1)}Y_{k_l}^{(2)}\right)B_{k_m}^{(1)}B_{k_o}^{(2)} +e^{ik_o}B_{k_j}^{(1)}B_{k_l}^{(2)}\left(\Gamma_{0\;3}^{2\;1}Y_{k_m}^{(2)}Y_{k_o}^{(1)}+\Gamma_{0\;3}^{1\;2}Y_{k_m}^{(1)}Y_{k_0}^{(2)}\right) \right\}=0,
\lb{f33}
\eea
where
\beq
D^{(3,3)}(k_j,k_l,k_m,k_o)=\Gamma_{0\;3}^{3\;0}+\Gamma_{3\;0}^{0\;3}e^{i(k_j+k_l+k_m+k_o)}+\left(\Gamma_{3\;0}^{3\;0}+\Gamma_{0\;3}^{0\;3}-\Gamma_{3\;3}^{3\;3} \right)e^{i(k_m+k_o)}.
\lb{f34}
\eeq
Since any pair $B_{k_j}^{(1)}B_{k_l}^{(2)}$ ($k_j \ne k_l$) is expressed in terms of the pair $Y_{k_j}^{(1)}Y_{k_l}^{(2)}$, through \rf{f24p}, we rewrite the expressions \rf{f30}, \rf{f31} and \rf{f32} only in terms of the matrices $\{Y_k^{(Q)} \}$. Multiplying \rf{f30} by the symmetric function $\rho(k_j,k_l)\rho(k_l,k_m)\rho(k_m,k_j)$ we obtain
\bea
&&\sum_{j,l,m=1}^3 \left\{ D^{(3,Q)}(k_j,k_l,k_m)D_3(k_l,k_m)\rho(k_j,k_l)Y_{k_j}^{(Q)}Y_{k_l}^{(1)}Y_{k_m}^{(2)}\right. \nonumber \\
&& -e^{ik_m}\Gamma_{Q\;3}^{3\;Q}D_3(k_j,k_l)\rho(k_l,k_m) Y_{k_j}^{(1)}Y_{k_l}^{(2)}Y_{k_m}^{(Q)} \nonumber \\
&&\left.+\left( \Gamma_{0\;\;\;3}^{Q'\;Q}Y_{k_j}^{(Q)}Y_{k_l}^{(Q')}Y_{k_m}^{(Q)}+\Gamma_{0\;\;3}^{Q\;Q'}Y_{k_j}^{(Q)}Y_{k_l}^{(Q)}Y_{k_m}^{(Q')} \right)e^{ik_m}\rho(k_j,k_l)\rho(k_l,k_m) \right\}\rho(k_m,k_j)=0,
\nonumber \\
&&\sum_{j,l,m=1}^3 \left\{ D^{(Q,3)}(k_j,k_l,k_m)D_3(k_j,k_l)\rho(k_l,k_m)Y_{k_j}^{(1)}Y_{k_l}^{(2)}Y_{k_m}^{(Q)}\right. \nonumber \\
&& -e^{i(k_l+k_m)}\Gamma_{3\;Q}^{Q\;3}D_3(k_l,k_m)\rho(k_j,k_l)Y_{k_j}^{(Q)}Y_{k_l}^{(1)}Y_{k_m}^{(2)} \nonumber \\
&&\left. +\left( \Gamma_{3\;\;\;0}^{Q'\;Q}Y_{k_j}^{(Q')}Y_{k_l}^{(Q)}Y_{k_m}^{(Q)}+\Gamma_{3\;\;0}^{Q\;Q'}Y_{k_j}^{(Q)}Y_{k_l}^{(Q')}Y_{k_m}^{(Q)} \right)e^{i(k_l+k_m)}\rho(k_j,k_l)\rho(k_l,k_m) \right\}\rho(k_m,k_j)=0, \nonumber \\
\lb{f35}
\eea
where $(Q,Q')=(1,2)$ or $(2,1)$. Multiplying \rf{f33} by the symmetric combination $\rho(k_j,k_l)\rho(k_l,k_o)\rho(k_o,k_m)\rho(k_m,k_j)\rho(k_l,k_m)\rho(k_j,k_o)$ we obtain
\bea
&&\sum_{j,l,m,o=1}^4 \left\{ \left[ \left(D^{(3,3)}(k_j,k_l,k_m,k_o)D_3(k_j,k_l)+\Gamma_{3\;0}^{1\;2}e^{i(k_l+k_m+k_o)}\rho(k_j,k_l) \right)D_3(k_m,k_o)\right. \right.\nonumber \\
&&\left. +\Gamma_{0\;3}^{1\;2}e^{ik_o}D_3(k_j,k_l)\rho(k_m,k_o) \right] Y_{k_j}^{(1)}Y_{k_l}^{(2)}Y_{k_m}^{(1)}Y_{k_o}^{(2)} \nonumber \\
&&+\Gamma_{3\;0}^{2\;1}e^{i(k_l+k_m+k_o)}D_3(k_m,k_o)\rho(k_j,k_l)Y_{k_j}^{(2)}Y_{k_l}^{(1)}Y_{k_m}^{(1)}Y_{k_o}^{(2)} \nonumber \\
&&\left.+\Gamma_{0\;3}^{2\;1}e^{ik_o}D_3(k_j,k_l)\rho(k_m,k_o)Y_{k_j}^{(1)}Y_{k_l}^{(2)}Y_{k_m}^{(2)}Y_{k_o}^{(1)}  \right\}\rho(k_l,k_o)\rho(k_m,k_j)\rho(k_l,k_m)\rho(k_j,k_o)=0.
\lb{f37}
\eea

	In order to have a matrix product  ansatz for the Hamiltonian \rf{f1} the relation \rf{f35}-\rf{f37} should be consistent with the two words commutation relations \rf{f26}, for any values of $k_j \in C$ ($j=1,\ldots,4$). The successive use of \rf{f26} in \rf{f35}-\rf{f37} allow us to rewrite the left side of these equations as a polynomial on the variables $e^{ikj}$ ($j=1,\ldots,4$). Since we do not want, on this level, to fix the spectral parameters $\{k_j \}$, we should impose that all the coefficients of these polynomials are zero. This will give further constraints on the coupling constant $\Gamma_{n\;o}^{l\;m}$ besides \rf{f12} and those imposed by the Yang-Baxter relations \rf{f27}.

	It will not occur any additional constraint for $n>4$. Although we did not consider here the problem of searching all the possible solutions of the general Hamiltonian satisfying \rf{f27} and \rf{f35}-\rf{f37}, we verified that those equations are satisfied by several know exact integrable chains. The solution where  
\bea
&\Gamma_{\beta\;\alpha}^{\alpha\;\beta}=1, \;\;\; \Gamma_{\alpha\;\beta}^{\alpha\;\beta}={\mbox{sign}}(\alpha - \beta)\sinh(\gamma)-\epsilon_0 \cosh(\gamma) \;\;\; (\alpha \ne \beta) \nonumber \\
&\Gamma_{\alpha\;\alpha}^{\alpha\;\alpha}=(\epsilon_{\alpha}-\epsilon_0 )\cosh(\gamma), \;\;\; (\alpha, \beta=0,1,2,3),
\lb{f38}
\eea
where $\epsilon_0, \epsilon_1, \epsilon_2, \epsilon_3=\pm 1$ and $\gamma$ a free complex parameter give us the anisotropic spin-$\frac{3}{2}$ Perk-Schultz model. The fermionic version, obtained by a Jordan-Wigner transformation, of the Hamiltonian \rf{f1} with the choices \rf{f38} and with $\epsilon_0=-\epsilon_1=-\epsilon_2=\epsilon_3=1$ is the anisotropic version of the Essler-Korepin-Schoutens model. 

        The solution where
\bea
&\Gamma_{\gamma\;\nu}^{\alpha\;\beta}=\delta_{\alpha+\beta, \gamma+\nu} \;\;\; (\alpha \ne \beta,\; \gamma \ne \nu), \nonumber \\
&\Gamma_{3\;0}^{0\;3}=\Gamma_{0\;3}^{3\;0}=0, \;\;\; \Gamma_{\alpha\;3}^{\alpha\;3}=\Gamma_{3\;\alpha}^{3\;\alpha}=-\frac{U}{2}\left(1+\delta_{\alpha,3} \right),
\lb{f39}
\eea
with $U$ a free parameter, give us after a Jordan-Wigner transformation, the standard Hubbard model with Coulomb on site interaction $U$ \cite{lieb}. The solutions where the non-zero coupling constants are
\bea
&\Gamma_{0\;\alpha}^{\alpha\;0}=\Gamma_{\alpha\;0}^{0\;\alpha}=1 \;\;\; (\alpha \ne 0), \;\;\; \Gamma_{1\;3}^{3\;1}=\Gamma_{3\;1}^{1\;3}=\Gamma_{2\;3}^{3\;2}=\Gamma_{3\;2}^{2\;3}=\epsilon, \Gamma_{3\;1}^{3\;1}=\Gamma_{3\;0}^{3\;0}+\Gamma_{1\;2}^{1\;2},\nonumber \\
&\Gamma_{1\;2}^{3\;0}=\Gamma_{3\;0}^{1\;2}=\epsilon\Gamma_{0\;3}^{2\;1}=\epsilon\Gamma_{2\;1}^{0\;3}=\Gamma_{0\;3}^{1\;2}=\Gamma_{1\;2}^{0\;3}=\epsilon\Gamma_{2\;1}^{3\;0}=\epsilon\Gamma_{3\;0}^{2\;1}=\sin(\theta) \nonumber \\
&\Gamma_{1\;2}^{2\;1}=\Gamma_{2\;1}^{1\;2}=-\epsilon\Gamma_{2\;0}^{0\;2}=-\epsilon\Gamma_{0\;2}^{2\;0}=-\frac{\epsilon}{2}\Gamma_{3\;0}^{3\;0}=\Gamma_{2\;1}^{2\;1}e^{2\eta}=\Gamma_{1\;2}^{1\;2}e^{-2\eta}=\cos(\theta) \nonumber \\
&\Gamma_{3\;2}^{3\;2}=\Gamma_{3\;0}^{3\;0}+\Gamma_{2\;1}^{2\;1}, \;\;\; \Gamma_{1\;3}^{1\;3}=\Gamma_{1\;2}^{1\;2}, \;\;\; \Gamma_{2\;3}^{2\;3}=\Gamma_{2\;1}^{2\;1}, \;\;\; \Gamma_{3\;3}^{3\;3}=\Gamma_{3\;0}^{3\;0}+\Gamma_{2\;1}^{2\;1}+\Gamma_{1\;2}^{1\;2}
\lb{f40}
\eea
and
\bea
&\Gamma_{0\;\alpha}^{\alpha\;0}=\Gamma_{\alpha\;0}^{0\;\alpha}=1 \;\;\; (\alpha \ne 0), \;\;\; \Gamma_{1\;3}^{3\;1}=\Gamma_{3\;1}^{1\;3}=\Gamma_{2\;3}^{3\;2}=\Gamma_{3\;2}^{2\;3}=\epsilon, \nonumber \\
&\Gamma_{1\;2}^{3\;0}=\Gamma_{3\;0}^{1\;2}=\epsilon\Gamma_{0\;3}^{2\;1}=\epsilon\Gamma_{2\;1}^{0\;3}=\epsilon e^{2\eta}\Gamma_{0\;3}^{1\;2}=\epsilon e^{2\eta}\Gamma_{1\;2}^{0\;3}=e^{-2\eta}\Gamma_{2\;1}^{3\;0}=e^{-2\eta}\Gamma_{3\;0}^{2\;1}=\sin(\theta) \nonumber \\
&\Gamma_{1\;2}^{2\;1}=\Gamma_{2\;1}^{1\;2}=-\epsilon\Gamma_{2\;0}^{0\;2}=-\epsilon\Gamma_{0\;2}^{2\;0}=\Gamma_{2\;1}^{2\;1}e^{2\eta}=\Gamma_{1\;2}^{1\;2}e^{-2\eta}=\cos(\theta) \nonumber \\
&\Gamma_{3\;0}^{3\;0}=2\Gamma_{0\;2}^{2\;0}+\sin^2(\theta)\frac{\left(e^{\eta}-\epsilon e^{-\eta} \right)^2}{\cos(\theta)}, \;\;\; \Gamma_{3\;1}^{3\;1}=\Gamma_{3\;2}^{3\;2}=\Gamma_{3\;0}^{3\;0}+\Gamma_{1\;2}^{1\;2}, \nonumber \\
&\Gamma_{1\;3}^{1\;3}=\Gamma_{2\;3}^{2\;3}=\Gamma_{2\;1}^{2\;1}, \;\;\; \Gamma_{3\;3}^{3\;3}=\Gamma_{3\;0}^{3\;0}+\Gamma_{2\;1}^{2\;1}+\Gamma_{1\;2}^{1\;2},
\lb{f41}
\eea
where $\epsilon=\pm 1$ and $\theta$, $\eta$ are free parameters, give us the two-parameter integrable models introduced in \cite{alcbar2}. It is interesting to mention that these two models contain as special cases the Hubbard model, the Essler-Korepin-Schoutens model, as well as the $q$-deformation of the extended Hubbard models introduced in \cite{ramos,gould}.

	In all the cases the spectral parameters ($k_1,\ldots,k_n$) are going to be fixed by the cyclic property of the trace. The relations \rf{f18}, \rf{f15}, \rf{f24p} and \rf{f26} imply that any amplitude of the matrix product  ansatz \rf{f5}-\rf{f6} are proportional to $\mbox{Tr} \left[Y_{k_1}^{(Q_1)}\cdots Y_{k_n}^{(Q_n)}E^{L}\Omega_P\right]$. Successive applications of the commutation relations \rf{f15}, \rf{f26} and \rf{f6p} give us
\bea
&&\mbox{Tr} \left[Y_{k_1}^{(Q_1)}\cdots Y_{k_n}^{(Q_n)}E^{L}\Omega_P\right]=e^{ik_jL}\sum_{Q_1',\ldots,Q_n'=1}^2 \langle Q_1,\ldots,Q_n|{\cal{T}}|Q_1',\ldots,Q_n'\rangle \nonumber \\
&&\times  \mbox{Tr} \left[Y_{k_1}^{(Q_1')}\cdots Y_{k_n}^{(Q_n')}E^{L}\Omega_P\right],
\lb{f42}
\eea
where we have used the identity
\beq
\sum_{Q_j'',Q_{j+1}''=1}^2 S_{Q_j'\;Q_j''}^{Q_j\;\;Q_{j+1}''}(k_j,k_j)=-1,
\lb{f43}
\eeq
and
\beq
\langle Q_1,\ldots,Q_n|{\cal{T}}|Q_1',\ldots,Q_n'\rangle = \sum_{Q_1'',\ldots,Q_n''=1}^2\prod_{i=1}^nS_{Q_i'\;Q_i''}^{Q_i\;\;Q_{i+1}''}(k_i,k_j),
\lb{f44}
\eeq
is the $2^n \times 2^n$-dimensional transfer matrix of an inhomogeneous $6$-vertex model (inhomogeneities $\{k_j \}$ along the vertical) with Boltzmann weights given by \rf{f15p} and \rf{f24}. The vertex model is defined on cylinder of transversal perimeter $n$ and periodic boundary condition
\beq
S_{Q_n'\;Q_n''}^{Q_n\;\;Q_{n+1}''}(k_n,k)=S_{Q_n'\;Q_n''}^{Q_n\;\;Q_1''}(k_n,k).
\lb{f45}
\eeq
The relation \rf{f42} fix the values of the spectral parameters $\{k_j \}$ as the solution of the equation
\beq
e^{-ik_jL}=\Lambda(k_j,\{k_l \}) \;\;\; (j=1,\ldots,n),
\lb{f46}
\eeq
where $\Lambda(k_j,\{k_l \})$ are the eigenvalues of the transfer matrix \rf{f44}. In order to complete our solution we need to evaluate the eigenvalue of the inhomogeneous transfer matrix \rf{f44}. We are in a similar situation as in section $4$, except that now the inhomogeneous vertex model is defined on a periodic lattice. The eigenvalues $\Lambda(k_j,\{k_l \})$ is obtained by setting in \rf{e130} and \rf{e131} $\phi(s)=1$. The spectral parameters $k_1,\ldots,k_n$ are then fixed by the equation
\beq
e^{-ik_jL}=\prod_{l=1}^{n}S_{2\;2}^{2\;2}(k_l,k_j)\prod_{l=1}^{n_1}\frac{S_{2\;2}^{2\;2}(k_j,k_l^{(1)})}{S_{2\;1}^{2\;1}(k_j,k_l^{(1)})},
\lb{f47}
\eeq
where the unknown parameters $k_l^{(1)}$ ($l=1,\ldots,n$) are fixed by
\beq
\prod_{l=1}^{n}\frac{S_{2\;1}^{2\;1}(k_l,k_j^{(1)})}{S_{2\;2}^{2\;2}(k_l,k_j^{(1)})}=\prod_{l=1}^{n_1}\frac{S_{2\;2}^{2\;2}(k_j^{(1)},k_l^{(1)})}{S_{1\;1}^{1\;1}(k_l^{(1)},k_j^{(1)})}\frac{S_{2\;1}^{2\;1}(k_l^{(1)},k_j^{(1)})}{S_{2\;1}^{2\;1}(k_j^{(1)},k_l^{(1)})},
\lb{f48}
\eeq
for $j=1,\ldots,n_1$.


\section{ The generalized XXZ chain in an open chain  }

        Differently from the former sections where the quantum chains are all defined on periodic lattices we consider here the formulation of the matrix product  ansatz in an open chain. We consider the generalized XXZ chain defined in section $2$ in an open lattice with diagonal $z$-magnetic fields acting only at the surface points. The Hamiltonian we want to solve is given by 
\beq
H_s=-{\cal{P}}_s \left\{ \frac{1}{2} \sum_{i=1}^{L-1} \left( \sigma_i^x \sigma_{i+1}^x+\sigma_i^y \sigma_{i+1}^y+\Delta\sigma_i^z \sigma_{i+1}^z \right) + h_L \sigma_1^z +h_R \sigma_L^z \right\} {\cal{P}}_s,
\lb{g1}
\eeq
where $\sigma^x$, $\sigma^y$, $\sigma^z$ are spin-$\frac{1}{2}$ Pauli matrices, $\Delta$ is the anisotropy and $h_L$, $h_R$ are the magnetic fields acting at the end points $1$ and $L$, respectively. The projector ${\cal{P}}_s$, as in section $2$, projects out from the associated Hilbert space the configurations where any two up spins are at distances smaller than $s$ (s=1,2,\ldots). The choice $h_1=h_L=0$ corresponds to the free boundary case.

        The exact solution of \rf{g1} for the case $s=1$ was obtained through the coordinate Bethe  ansatz in \cite{ab3q} and through the quantum inverse scattering method in \cite{sklyanin}. It is also interesting to mention that, in the case where $s=1$ the Hamiltonian \rf{g1} is SU($2$)$_q$-invariant if the anisotropy and surface fields are related by $\Delta=(q+1/q)/2$, $h_L=q$, $h_R=1/q$.

        Since the Hamiltonian \rf{g1} commutes with the total spin operator $S^z=\sum_{i=1}^L\sigma_i^z$ the number of up spins $n$ is a good quantum number. We want to solve the eigenvalue equation
\beq
H|\Psi_n\rangle=\varepsilon_n|\Psi_n\rangle
\lb{g2}
\eeq
where
\beq
|\Psi_n\rangle=\sum f(x_1,\ldots,x_n)|x_1,\ldots,x_n\rangle.
\lb{g3}
\eeq
Here $x_1,\ldots,x_n$ denote the configurations of the up spins on the chain, and the summation extends over all sets of $n$ increasing integers satisfying
\beq
x_1 \ge 1, \;\;\; x_n \le L, \;\;\; x_{i+1} \ge x_i +s, \;\;\; (i=1,\ldots,n-1).
\lb{g4}
\eeq
As in section $2$ in order to formulate our matrix product  ansatz we associate the matrices $E$ and $A$ to the sites occupied by down and up spins, respectively \footnote{Differently from the solution on the periodic lattice (see \rf{e4} in section $2$) in the present case it is not necessary to define the matrices $A$ with the superscript $s$.}. Our  ansatz asserts that any amplitude in \rf{g3} is given by
\beq
f(x_1,\ldots,x_n)=E^{x_1-1}AE^{x_2-x_1-1}A\cdots AE^{x_n-x_{n-1}-1}AE^{L-x_n}.
\lb{g5}
\eeq
Actually $E$ and $A$ are abstract operators with an associative product. A well defined eigenfunction \rf{g3} is obtained, apart from a normalization, if all the amplitudes are related uniquely. In order to obtain the solutions through the  ansatz \rf{g5} let us consider initially the cases of small values of $n$.

{\it {\bf  n = 1.}}
For one up spin the eigenvalue equation \rf{g2} give us three types of relations depending if the corresponding configuration has the up spin at $x=2,\ldots,L-1$ or at the boundaries $x=1$ and $x=L$, namely,
\bea
&\varepsilon_1 E^{x-1}AE^{L-x}=-E^{x-1}\left(E^{-1}AE+EAE^{-1} \right)E^{L-x} \nonumber \\
&-\frac{1}{2}\left[\left(L-5 \right)\Delta+h_L+h_R \right]E^{x-1}AE^{L-x}, \;\;\;\; x=2,\ldots,L-1
\lb{g6}
\eea
\beq
\varepsilon_1AE^{L-1}=-EAE^{L-2}-\frac{1}{2}\left[\left(L-3 \right)\Delta-h_L+h_R \right]AE^{L-1}
\lb{g7}
\eeq
\beq
\varepsilon_1E^{L-1}A=-E^{L-2}AE-\frac{1}{2}\left[\left(L-3 \right)\Delta+h_L-h_R \right]E^{L-1}A.
\lb{g8}
\eeq
The solution of all these equations is obtained by identifying the matrix $A$ as composed by two other matrices $B_k$, $C_k$ depending on a single spectral parameter $k$, i. e.,
\beq
A=(B_k-C_k)E^{2-s},
\lb{g9}
\eeq
with the following commutation relation with the matrix $E$
\beq
EB_k=e^{ik}B_kE, \;\;\; EC_k=e^{-ik}C_kE.
\lb{g10}
\eeq 
Substituting \rf{g9} in \rf{g6} and using \rf{g10} we obtain the energy in terms of the spectral parameter $k$,
\beq
\varepsilon_1=-2\cos(k)-\frac{1}{2}\left[\left(L-5 \right)\Delta+h_L+h_R \right].
\lb{g11}
\eeq
Inserting \rf{g11}, \rf{g9} in \rf{g7}, \rf{g8} and using \rf{g10} we obtain the following algebraic relations
\beq
\alpha(k)e^{-ik}B_k -\alpha(-k)e^{ik}C_k=0
\lb{g12}
\eeq
\beq
\beta(-k)e^{ik}B_k -\beta(k)e^{-ik}C_k=0,
\lb{g13}
\eeq
where $\alpha(k)$ and $\beta(k)$ are given by
\beq
\alpha(k)=1+(h_L-\Delta)e^{ik}, \;\;\; \beta(k)=[1+(h_R-\Delta)e^{ik}]e^{-ik(L+1)}.
\lb{g14}
\eeq
The compatibility of the relations \rf{g12} and \rf{g13} fix the spectral parameter
\beq
\frac{\alpha(k)\beta(k)}{\alpha(-k)\beta(-k)}=1.
\lb{g15}
\eeq

{\it {\bf  n = 2.}}
 The eigenvalue equation produces now fours types of relations depending on the relative location $x_1$, $x_2$ of the up spins. The amplitudes related to the configurations where $x_1>1$, $L>x_2>x_1+s$ give us
\bea
\epsilon_2 E^{x_1-1}AE^{x_2-x_1-1}AE^{L-x_2}&=&-E^{x_1-1}\left(E^{-1}AE+EAE^{-1} \right)E^{x_2-x_1-1}E^{L-x_2} \nonumber \\
&&-E^{x_1-1}AE^{x_2-x_1-1}\left(E^{-1}AE+EAE^{-1} \right)E^{L-x_2} \nonumber \\
&&-\frac{1}{2}\left[(L-9)\Delta+h_L+h_{R}\right]E^{x_1-1}AE^{x_2-x_1-1}AE^{L-x_2},
\lb{g17}
\eea
while the configurations where the particles are at the matching conditions $x_1>1$, $L>x_2=x_1+1$ produce the relation
\bea
&\epsilon_2 E^{x_1-1}AE^{s-1}AE^{L-x_1-s}=-E^{x_1-2}AE^sAE^{L-x_1-s} \nonumber \\
&-E^{x_1-1}AE^sAE^{L-x-s-1}-\frac{1}{2}\left[(L-5)\Delta+h_{L}+h_{R}\right]E^{x_1-1}AE^{s-1}AE^{L-x-s}.
\lb{g18}
\eea
Lastly the amplitudes where one of the particles are at the end points give us the following relations. For $x_1=1$, $L>x_2>x_1+s$,
\bea
&\epsilon_2 AE^{x_2-2}AE^{L-x_2}=-EAE^{x_2-3}AE^{L-x_2}-AE^{x_2-3}AE^{L-x_2+1} \nonumber \\
&-AE^{x_2-1}AE^{L-x_2-1}-\frac{1}{2}\left[(L-7)\Delta-h_{L}+h_{R}\right]AE^{x_2-2}AE^{L-x_2},
\lb{g19}
\eea
and for $1<x_1<L-s$, $x_2=L$
\bea
&\epsilon_2 E^{x_1-1}AE^{L-x_1-1}A=-E^{x_1-2}AE^{L-x_1}A-E^{x_1}AE^{L-x_1-2}A \nonumber \\
&-E^{x_1-1}AE^{L-x_1-2}AE-\frac{1}{2}\left[(L-7)\Delta+h_{L}-h_{R}\right]E^{x_1-1}AE^{L-x_1-1}A.
\lb{g20}
\eea
The relations coming from the amplitudes where the particles are located at ($x_1=1$, $x_2=1+s$) or ($x_1=L-s$, $x_2=L$) are satisfied by the solutions of \rf{g17}-\rf{g20}. The solution of \rf{g17}-\rf{g20} is obtained by a generalization of \rf{g9}, where we identify the matrix $A$ as composed by $n=2$ pairs of spectral parameter matrices $\{B_{k_j},C_{k_j}\}$,
\beq
A=\sum_{j=1}^{n}\left(B_{k_j}-C_{k_j} \right)E^{2-s},
\lb{g21}
\eeq
obeying the following commutation relations
\bea
&&EB_{k_j}=e^{ik_j}B_{k_j}E, \;\;\;  EC_{k_j}=e^{-ik_j}C_{k_j}E \nonumber \\
&&B_{k_j}^2=C_{k_j}^2=B_{k_j}C_{k_j}=C_{k_j}B_{k_j}=0  \;\;\; j=1,\ldots,n.
\lb{g22}
\eea
Inserting \rf{g21} into \rf{g17} we obtain the energy $\epsilon_2$ in terms of the unknown spectral parameters $k_1$, $k_2$;
\beq
\epsilon_n=-2\sum_{j=1}^n \cos(k_j)-\frac{1}{2}\left[ (L-1-4n)\Delta+h_{L}+h_{R}\right],
\lb{g23}
\eeq
where $n=2$. The ``bulk'' relations \rf{g18} give us, after using \rf{g21} and \rf{g22}, the following algebraic relations among the matrices $\{B_{k_j},C_{k_j}\}$:
\bea
&&a(k_l,k_j)B_{k_j}B_{k_l}+a(k_j,k_l)B_{k_l}B_{k_j}=0   \nonumber \\
&&a(-k_l,-k_j)C_{k_j}C_{k_l}+a(-k_j,-k_l)C_{k_l}C_{k_j}=0   \nonumber \\
&&a(-k_l,k_j)B_{k_j}C_{k_l}+a(k_j,-k_l)C_{k_l}B_{k_j}=0  \;\;\; j,l=1,\ldots,n \;\;\; (j\ne l)
\lb{g26}
\eea
with $n=2$ and
\beq
a(k,k')=1-2\Delta e^{-ik'}+e^{-i\left(k+k'\right)}.
\lb{g27}
\eeq
Using \rf{g21} and \rf{g22} in \rf{g19} and \rf{g20} give us the following additional relations
\bea
&&\alpha(k_j)e^{-ik_j}B_{k_j}B_{k_l}-\alpha(-k_j)e^{ik_j}C_{k_j}B_{k_l}=0  \lb{g28} \\
&&\alpha(-k_j)e^{ik_j}C_{k_j}C_{k_l}-\alpha(k_j)e^{-ik_j}B_{k_j}C_{k_l}=0   \lb{g29} \\
&&\beta(-k_l)e^{-ik_ls}B_{k_j}B_{k_l}-\beta(k_l)e^{ik_ls}B_{k_j}C_{k_l}=0  \lb{g30}  \\
&&\beta(k_l)e^{ik_ls}C_{k_j}C_{k_l}-\beta(-k_l)e^{-ik_ls}C_{k_j}B_{k_l}=0, \;\;\; j.l=1,\ldots,n \;\;\; (j\ne l) \lb{g31}
\eea
with $n=2$ and
\beq
\alpha(k)=1+(h_L-\Delta)e^{ik}, \;\;\; \beta(k)=[1+(h_R-\Delta)e^{ik}]e^{-ik(L+1)}.
\lb{g32}
\eeq
The up to now free spectral parameters $k_1$ and $k_2$ are going to be fixed by imposing the compatibility of the algebraic relations \rf{g26}, (\rf{g28}-\rf{g31}). Using successively \rf{g28}, \rf{g26}, \rf{g30} and \rf{g26} we obtain
\beq
\frac{\alpha(k_j)\beta(k_j)}{\alpha(-k_j)\beta(-k_j)}e^{2ik_j(s-1)}=\frac{{\cal B}(-k_j,k_l)}{{\cal B}(k_j,k_l)}, \;\;\; j=1,2, \;\;\; l \ne j
\lb{g33}
\eeq
where
\beq
{\cal B}(k,k')=a(k,k')a(k',-k).
\lb{g34}
\eeq

{\it {\bf  General n.}}
The eigenvalue equation applied to the components corresponding to the configurations where there exist no collisions ($x_{i+1}>x_i+s$; $i=1,\ldots,n-1$) produces a generalization of the relation \rf{g17} that is solved by identifying, as in \rf{g21}, the matrix $A$ as composed by $n$ pairs of spectral parameter dependent matrices $\{B_k,C_k \}$ satisfying the algebraic relations \rf{g22}. In terms of the spectral parameters $k_1,\ldots,k_n$ the energy is given by \rf{g23}. The eigenvalue equation \rf{g2} when applied to the components related to the configurations where two particles at $x_i$ and $x_{i+1}$ are at the ``colliding'' positions $L>x_{i+1}=x_i+s>s$ give us the relations \rf{g26} for $j \ne l$, $j,l=1,\ldots,n$. The configurations where we have a particle at the end points $x_1=1$ or $x_n=L$ give us the additional relations
\bea
&&\left[B_{k_{i_1}}B_{k_{i_2}}\alpha(k_{i_1})e^{-ik_{i_1}}-C_{k_{i_1}}B_{k_{i_2}}\alpha(-k_{i_1})e^{ik_{i_1}} \right]X_{k_{i_3}}\cdots X_{k_{i_n}}=0, \nonumber \\
&&\left[C_{k_{i_1}}C_{k_{i_2}}\alpha(-k_{i_1})e^{ ik_{i_1}}-B_{k_{i_1}}C_{k_{i_2}}\alpha(k_{i_1})e^{-ik_{i_1}} \right]X_{k_{i_3}}\cdots X_{k_{i_n}}=0, \nonumber \\
&&X_{k_{i_1}}\cdots X_{k_{i_{n-2}}}\left[B_{k_{i_{n-1}}}B_{k_{i_n}}\beta(-k_{i_n})e^{ik_{i_n}[(n-1)(1-s)-1]}-B_{k_{i_{n-1}}}C_{k_{i_n}}\beta(k_{i_n})e^{-ik_{i_n}[(n-1)(1-s)-1]} \right]= \nonumber \\
&&0, \nonumber \\
&&X_{k_{i_1}}\cdots X_{k_{i_{n-2}}}\left[C_{k_{i_{n-1}}}C_{k_{i_n}}\beta(k_{i_n})e^{-ik_{i_n}[(n-1)(1-s)-1]}-C_{k_{i_{n-1}}}B_{k_{i_n}}\beta(-k_{i_n})e^{ik_{i_n}[(n-1)(1-s)-1]} \right]= \nonumber \\
&&0, \lb{g35}
\eea
where $i_1,\ldots,i_n$ is an arbitrary permutation of the integers $1,2,\ldots,n$, and $X_{k_j}$ denote a matrix $B_{k_j}$ or $C_{k_j}$. It is interesting to observe that while the algebraic relations \rf{g26} only relate two product of two matrix the relations \rf{g35} relate the product of $n$ matrices.

        The matrix product  ansatz \rf{g5} works only if all the amplitudes of the eigenfunction \rf{g3} are uniquely related. In fact the algebraic relations \rf{g22}, \rf{g26} and \rf{g35} enable us to show that any amplitude given by the  ansatz \rf{g5} is proportional to the matrix product $B_{k_1}\cdots B_{k_n}E^{L-n}$. The spectral parameters $k_1,\ldots,k_n$ are fixed by imposing the compatibility of the algebraic relations \rf{g26} and \rf{g35}. For any $j=1,\ldots,n$ we have
\bea
&&B_{k_1}\cdots B_{k_j}\cdots B_{k_n}=\left[\prod_{l=j+1}^n \frac{a(k_j,k_l)}{a(k_l,k_j)} \right]\frac{\beta(k_j)}{\beta(-k_j)}e^{-2ik_j[(n-1)(1-s)-1]}B_{k_1}\cdots B_{k_{j-1}}B_{k_{j+1}}\cdots B_{k_n}C_{k_j} \nonumber \\
&&=e^{-2ik_j(n-1)(1-s)}\frac{\alpha(k_j)\beta(k_j)}{\alpha(-k_j)\beta(-k_j)}\prod_{l=1, \; l \ne j}^n \frac{a(k_j,k_l)a(k_l,-k_j)}{a(k_l,k_j)a(-k_j,k_l)}B_{k_1}\cdots B_{k_j}\cdots B_{k_n}
\lb{g39}
\eea
that give us
\beq
\frac{\alpha(k_j)\beta(k_j)}{\alpha(-k_j)\beta(-k_j)}e^{2ik_j(n-1)(s+1)}=\prod_{l=1, \; l \ne j}^n \frac{{\cal B}(-k_j,k_l)}{{\cal B}(k_j,k_l)}, \;\;\; j=1,\ldots,n.
\lb{g40}
\eeq
The eigenvalues are given by \rf{g23} with the spectral parameters obtained by the solutions of \rf{g40}. The equations \rf{g23} and \rf{g40} for $s=1$ coincides with the known result derived in \cite{ab3q} through the coordinate Bethe  ansatz.


\section{ Conclusions and generalizations }

        Formulations of a matrix product  ansatz were introduced along the years in order to describe the ground state wave function of some special quantum chains. In general only the ground state wave function of these models is described by this  ansatz and the quantum chain is not exact integrable. An exception happens in the formulation named dynamical matrix product  ansatz where the matrices defining the  ansatz are time dependent. In this case the full eigenspectra  of some exact integrable quantum chains related to stochastic model are derived. We have shown in this paper that a huge family of exact integrable quantum chains, normally solved through the Bethe  ansatz can also be solved by an appropriate matrix product  ansatz. In our formulation, independently if the quantum chain is related or not to a stochastic model, the matrices are time independent. Differently from the Bethe  ansatz where the amplitudes of eigenfunctions are given by combinations of plane waves, in the matrix product  ansatz these amplitudes are given by a product of matrices.

        A necessary condition for the integrability of the model through the matrix product  ansatz is the existence of an associative product among the matrices defining the  ansatz. In addition, since the amplitudes of the eigenfunctions are related to these matrix products, the algebraic properties of the matrices should provide a single relation among any two matrix products appearing in the { ansatz. These algebraic relations are obtained by imposing the eigenvalue equation and depend on the coupling constants defining the quantum chain.

        We have shown the formulation of the matrix product  ansatz for two classes of models. Models with a single global conservation law (U($1$) symmetry) like the XXZ chain (see sections $2$ and $3$), the spin-$1$ Fateev-Zamolodchikov model, and models with two conservation laws (U($1$)$\otimes$U($1$) symmetry) like the spin-$1$ Perk-Schultz models, the Hubbard models as well the other models presented in sections $4$ and $5$.

        Let us discuss initially the models with a single conservation law. The associativity of the algebra, in this case, is immediate since the structure constants defining the algebraic relations among the matrices in the  ansatz are complex constant numbers. In the case of the XXZ chain the algebraic relations merged in the sector with two particles are enough to ensure the exact integrability of the quantum chain (see section $2$). For the spin-$1$ models there appear additional relations involving the product of $3$ and $4$ matrices since the coupling constants defining the Hamiltonian connect up to four particles at nearest neighbor sites. The generalization of the matrix product  ansatz for higher spin models ($s>1$), although we did not consider in this paper follows straightforwardly. For example for spin $s=3/2$ models we should relate to the spins $-3/2$, $-1/2$, $1/2$ and $3/2$, in $S^z$-basis, the matrices $E$, $A$, $BE^{-1}B$ and $CE^{-1}CE^{-1}C$, respectively. It will appear in this case algebraic relations involving the product of $3$, $4$, $5$ and $6$ matrices.

        In the case of models with two global conservation laws (sections $4$ and $5$) the structure constants defining the algebraic relations among the product of two matrices are also matrices ($S$-matrix) and the associativity condition is equivalent to the famous Yang-Baxter relations for the $S$-matrices. In the case of the spin-$1$ Perk-Schultz model (see section $4$) the associativity of the algebra is enough to ensure the exact integrability of the quantum chain, however in the case of the Hubbard model, as well the other quantum chains presented in section $5$, there appear additional relations among the product of three and four matrices.

        Generalizations of the present matrix product  ansatz to the cases where we have three or more conservation laws follows straightforwardly. As we have shown along this paper (see sections $2$, $3$, $4$ and $6$) our formulation of the matrix product  ansatz allows the extension of several exact integrable models by including arbitrary hard-core effects, without destroying their exact integrability.
        Also it is interesting to mention that in the cases of exact integrable Hamiltonians associated to stochastic models, as in \cite{alcrit1,reviewshutz}, since we can write all eigenfunctions in a matrix product formulation, our results imply that we can equivalently write at any time the probability distribution of the model in terms of a time-dependent matrix product  ansatz as happens in the formulation of the dynamical matrix product  ansatz.

        Except in section $6$, all the quantum chains considered in this paper are defined on a periodic lattice and the eigenenergies are fixed by the cyclic property of the trace of the product of matrices appearing in the  ansatz. In section $6$ we show how to formulate the matrix product  ansatz in the case where the quantum chains are defined on open lattices. We derived the solution of the XXZ chain with magnetic fields at the endpoints of the lattice. The formulation of a matrix product  ansatz for the other models of sections $3$, $4$ and $5$ with open boundary conditions that preserve their global symmetry is also possible.

        Since the exact integrable chains considered in this paper share the same eigenfunctions with a related two dimensional vertex model the matrix product  ansatz we formulated provide also a solution for these classical models.
        A quite interesting problem for the future concerns the formulation of the matrix product  ansatz for the quantum chains with no global conservation law like the XYZ model, the $8$ vertex model or the case where the quantum chains are defined on open lattices with non-diagonal boundary fields.

In conclusion our results induce us  to conjecture that a matrix 
product {\it anstaz}, along the lines presented in this paper, can be formulated for 
any exact integrable quantum chain. The importance of this  ansatz for the 
future, as shown in this paper, remains on its simplicity, allowing quite
simple generalizations and the formulation of knew exact integrable models.

\acknowledgements{ This work has been partially supported by FAPESP, CNPq and CAPES (Brazilian agencies).}
\newpage 

\end{document}